\documentclass[aps,prl,twocolumn,groupedaddress,amsmath,superscriptaddress]{revtex4-2}

\usepackage{amsfonts}
\usepackage{amsmath}
\usepackage[dvips]{graphicx}
\usepackage{color}
\usepackage{amsmath}
\usepackage{graphicx}
\usepackage{amssymb}
\usepackage{hyperref}
\usepackage{color}
\usepackage{mathrsfs}
\usepackage{isomath}
\usepackage{amsmath}
\usepackage{amsthm}
\usepackage{epstopdf}
\usepackage{txfonts}
\usepackage{dsfont}
\usepackage{float}
\usepackage{setspace}
\usepackage{multirow}
\usepackage{booktabs}
\usepackage{tabularx}
\usepackage[T1]{fontenc}
\usepackage{colortbl}
\usepackage{hhline}
\usepackage{xcolor}
\usepackage[normalem]{ulem}
\allowdisplaybreaks[4]

\definecolor{nblue}{rgb}{0.3,0.3,1.0}
\definecolor{ngreen}{rgb}{0.2,0.7,0.2}
\definecolor{nred}{rgb}{0.9,0.1,0}
\definecolor{nblack}{rgb}{0,0,0}

\newcommand{\beq}{\begin{equation}}
\newcommand{\eeq}{\end{equation}}
\newcommand{\bqa}{\begin{eqnarray}}
\newcommand{\eqa}{\end{eqnarray}}

\definecolor{nyellow}{rgb}{1.0,0.75,0.0}

\makeatletter

\usepackage{babel}

\begin{document}
	
	\title{Randomness versus Nonlocality in Multi-input and Multi-output Quantum Scenario}
	\author{Chao Zhang}
	\thanks{These authors contributed equally to this work.}
	\affiliation
	{CAS Key Laboratory of Quantum Information, University of Science and Technology of China, Hefei 230026, China}
	\affiliation{CAS Synergetic Innovation Center of Quantum Information $\&$ Quantum Physics, University of Science and Technology of China, Hefei 230026, China}
	\author{Yi Li}
	\thanks{These authors contributed equally to this work.}
	\affiliation{State Key Laboratory for Mesoscopic Physics, School of Physics, Frontiers Science Center for Nano-optoelectronics, $\&$ Collaborative Innovation Center of Quantum Matter, Peking University, Beijing 100871, China}
	\affiliation{Beijing Academy of Quantum Information Sciences, Beijing 100193, China}
	\author{Xiao-Min Hu}
	\thanks{These authors contributed equally to this work.}
	\affiliation
	{CAS Key Laboratory of Quantum Information, University of Science and Technology of China, Hefei 230026, China}
	\affiliation{CAS Synergetic Innovation Center of Quantum Information $\&$ Quantum Physics, University of Science and Technology of China, Hefei 230026, China}
	\author{Yu Xiang}
	\email{xiangy.phy@pku.edu.cn}
	\affiliation{State Key Laboratory for Mesoscopic Physics, School of Physics, Frontiers Science Center for Nano-optoelectronics, $\&$ Collaborative Innovation Center of Quantum Matter, Peking University, Beijing 100871, China}
	\affiliation{Collaborative Innovation Center of Extreme Optics, Shanxi University, Taiyuan, Shanxi 030006, China}
	\author{Chuan-Feng Li}
	\affiliation
	{CAS Key Laboratory of Quantum Information, University of Science and Technology of China, Hefei 230026, China}
	\affiliation{CAS Synergetic Innovation Center of Quantum Information $\&$ Quantum Physics, University of Science and Technology of China, Hefei 230026, China}
	\author{Guang-Can Guo}
	\affiliation
	{CAS Key Laboratory of Quantum Information, University of Science and Technology of China, Hefei 230026, China}
	\affiliation{CAS Synergetic Innovation Center of Quantum Information $\&$ Quantum Physics, University of Science and Technology of China, Hefei 230026, China}
	\author{Jordi Tura}
	\affiliation{$\langle aQa^L\rangle$ Applied Quantum Algorithms Leiden, The Netherlands}
	\affiliation{Instituut-Lorentz, Universiteit Leiden, P.O. Box 9506, 2300 RA Leiden, The Netherlands}
	\author{Qihuang Gong}
	\affiliation{State Key Laboratory for Mesoscopic Physics, School of Physics, Frontiers Science Center for Nano-optoelectronics, $\&$ Collaborative Innovation Center of Quantum Matter, Peking University, Beijing 100871, China}
	\affiliation{Collaborative Innovation Center of Extreme Optics, Shanxi University, Taiyuan, Shanxi 030006, China}
	\affiliation{Peking University Yangtze Delta Institute of Optoelectronics, Nantong, Jiangsu 226010, China}	
	\affiliation{Hefei National Laboratory, Hefei 230088, China}
	\author{Qiongyi He}
	\email{qiongyihe@pku.edu.cn}
	\affiliation{State Key Laboratory for Mesoscopic Physics, School of Physics, Frontiers Science Center for Nano-optoelectronics, $\&$ Collaborative Innovation Center of Quantum Matter, Peking University, Beijing 100871, China}
	\affiliation{Collaborative Innovation Center of Extreme Optics, Shanxi University, Taiyuan, Shanxi 030006, China}
	\affiliation{Peking University Yangtze Delta Institute of Optoelectronics, Nantong, Jiangsu 226010, China}	
	\affiliation{Hefei National Laboratory, Hefei 230088, China}
	\author{Bi-Heng Liu}
	\email{bhliu@ustc.edu.cn}
	\affiliation{CAS Key Laboratory of Quantum Information, University of Science and Technology of China, Hefei 230026, China}
	\affiliation{CAS Synergetic Innovation Center of Quantum Information $\&$ Quantum Physics, University of Science and Technology of China, Hefei 230026, China}
	
	\begin{abstract}
		Device-independent randomness certification based on Bell nonlocality does not require any assumptions about the devices and therefore provides adequate security. Great effort has been made to demonstrate that nonlocality is necessary for generating quantum randomness, but the minimal resource required for random number generation has not been clarified. Here we first prove and experimentally demonstrate that violating any two-input Bell inequality is both necessary and sufficient for certifying randomness, however, for the multi-input cases, this sufficiency ceases to apply, leading to certain states exhibiting Bell nonlocality without the capability to certify randomness. We examine two typical classes of Bell inequalities with multi-input and multi-output, the facet inequalities and Salavrakos-Augusiak-Tura-Wittek-Ac\'{i}n-Pironio Bell inequalities, in the high-dimensional photonic system, and observe the violation of the latter one can always certify randomness which is not true for the former. 
		The private randomness with a generation rate of $1.867\pm0.018$ bits per photon pair is obtained in the scenario of Salavrakos-Augusiak-Tura-Wittek-Ac\'{i}n-Pironio Bell inequalities with $3$-input and $4$-output. Our work unravels the internal connection between randomness and nonlocality, 
		and effectively enhances the performance of tasks such as device-independent random number generation.
	\end{abstract}
	\maketitle
	
	\textit{Introduction.---} Born's rule~\cite{born_1926} guarantees the \textit{intrinsic} randomness that arises independently of any lack of knowledge or stochastic behaviors,  establishing it as a fundamental axiom in quantum theory. Building on this phenomenon, true random numbers can be generated~\cite{rmp_2017_random} relying on a pure source and well-characterized measurement apparatuses, exemplified by the unpredictable outcomes produced by a single-photon source and a 50:50 beam splitter~\cite{thomas}. However, noise on the source or devices is inevitable in a real laboratory, providing attackers with opportunities to gather information about the measurement outcomes, thereby diminishing the security of these results. This issue can be addressed through a Bell inequality since its violation expresses nonlocality and thus the measurement outcomes can rule out the \textit{deterministic} local hidden-variable model~\cite{rmp_2014_nonlocality}, which enables the certification of randomness without any knowledge about the source and the devices~\cite{nature_2010_random}. Therefore, the randomness generation based on Bell nonlocality has been intensively studied in both theoretical~\cite{njp_2014_the,jpa_2014_the,prl_2015_acin,pra_2016_tamas,prl_2012_acin,prl_2022_roger,prr_2020_acin,pra_2023_chain,njp_2016_chain} and experimental domains~\cite{science_2018_jianweiwang,np_2021_ustc,nphy_2021_random,prl_2021_ustc,nature_2018_ustc}.
	
	In principle, two bits of randomness, the highest achievable randomness in a two-input and two-output scenario, can be achieved when a modified Clauser-Horne-Shimony-Holt (CHSH) inequality is maximally violated~\cite{jpa_2014_the}. 
	Nevertheless, owing to the unavoidable loss and decoherence encountered in distributing nonlocal resources, along with the stringent requirements of detection losses to carry out a loophole-free Bell test, only a modest violation of Bell inequality can be observed~\cite{prl_2013_belltest,nature_2018_ustc,nature_2013_belltest,prl_2022_belltest}.
	Therefore, nonlocal resources between two distant stations are scarce in realistic scenarios. Consequently, it is natural to show keen interest in the relation between nonlocality and certifiable randomness, which helps us determine the minimal quantum resource for randomness certification. 
	
	In the two-input and two-output scenario, where the behaviors can be reproduced by a two-qubit system, all the nonlocal behaviors tested from CHSH inequality contribute to the randomness certification~\cite{nature_2010_random,prl_2012_acin}, i.e., one can certify nonzero randomness once the CHSH inequality is violated, however slightly. Further extending our focus to the more general multi-input and multi-output (MIMO) scenario, it is known to aid in decreasing the threshold of detection efficiency~\cite{prl_2010_belltest,prl_2022_belltest}, revealing more entangled states~\cite{pra_2008_Vertesi,np_2010_howard}, raising the thresholds of the highest achievable randomness~\cite{prl_2015_acin,science_2018_jianweiwang}, but it faces difficulty in reproducing all behaviors in a simplified low-dimensional system~\cite{rmp_2014_nonlocality}. Usually, MIMO scenario can be achieved through high-dimensional systems which have many advantages in performing device-independent (DI) tasks~\cite{PhysRevLett.88.127902,doi:10.1126/sciadv.abc3847,PhysRevLett.123.170402} and can obtain larger violation of Bell inequalities~\cite{PhysRevLett.88.040404}. However, to the best of our knowledge, the question of whether a MIMO Bell inequality is still necessary and sufficient to witness randomness remains unresolved.
	
	
	In this work, we first certify randomness when a given Bell inequality with arbitrary inputs and outputs is violated and further explore the relationship between Bell nonlocality and randomness. We prove that for all Bell inequalities with two inputs, their violation is a necessary and sufficient condition for the Bell inequality-based randomness certification of the system. However, this statement no longer holds for those Bell inequalities with multiple inputs. With the ability to accurately quantify randomness, we experimentally generate randomness in high-dimensional systems when MIMO Bell inequality violations are observed.
	Basic examples including two typical MIMO Bell inequalities, the facet Bell inequalities~\cite{prl_2002_CGLMP,njp_2004_Immnn,pla_2008_I4422} and Salavrakos-Augusiak-Tura-Wittek-Ac\'{i}n-Pironio (SATWAP) inequalities~\cite{prl_2017_SATWAP}, are demonstrated. We find that with multiple inputs, there is a regime where the facet inequality is violated but, surprisingly, there is no randomness unless the violation is large enough. In contrast, the SATWAP inequality is more sensitive for this task, so it can act as a good witness for certifying randomness in a broader range of detection efficiency of $80.0\%$.
	
	
	\begin{figure*}[ht!]
		\centering
		\includegraphics[width=0.9\textwidth]{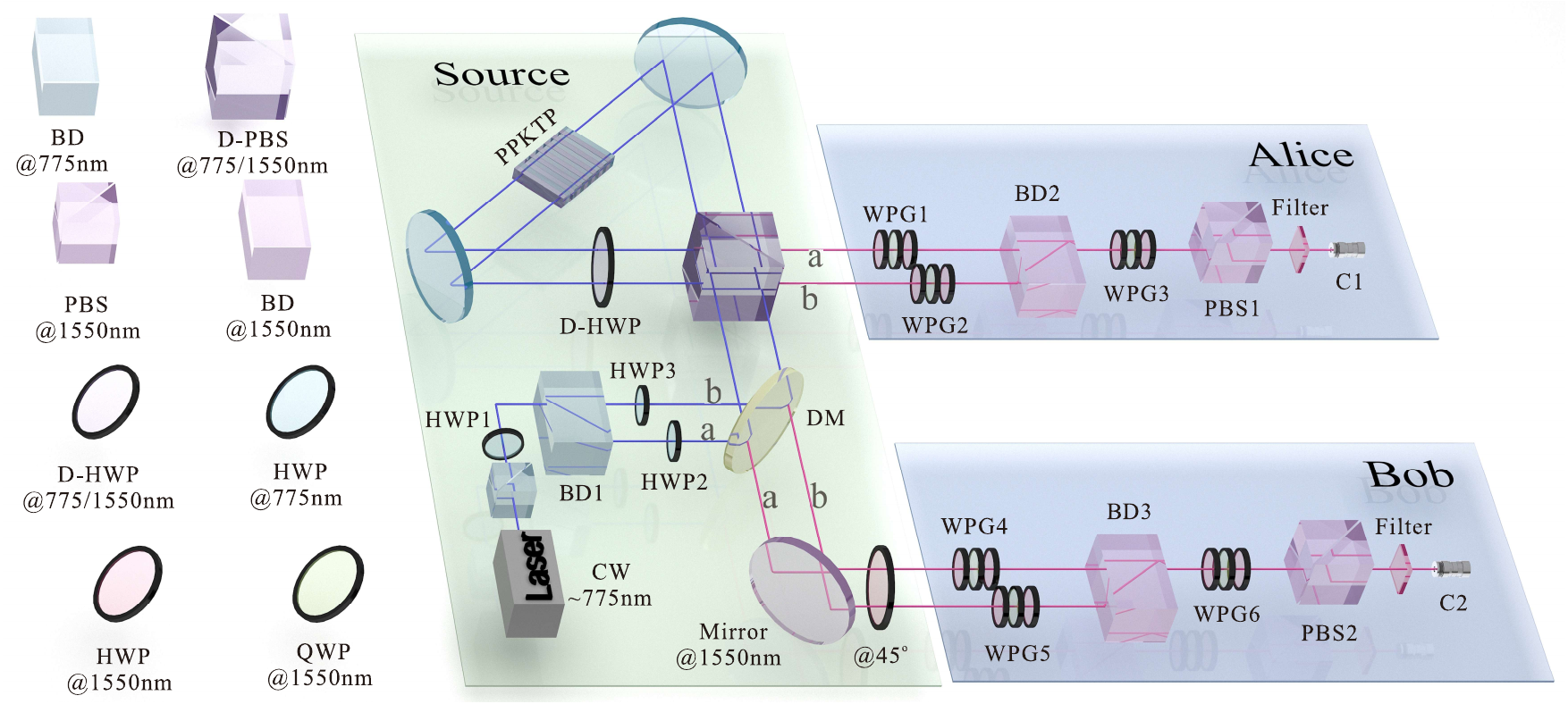}
		\vspace{-0.3cm}
		\caption{\textmd{Experiment setup.} A continuous-wave laser operating at a wavelength of 775 nm is utilized to pump the PPKTP crystal in a Sagnac structure. This process generates correlated photons with a wavelength of 1550 nm, exhibiting an entangled state represented as $|\Psi\rangle$. The parameters $\theta_1$, $\theta_2$, and $\theta_3$ can be individually controlled by HWP1-HWP3, enabling the preparation of two-, three-, and four-dimensional entangled states. The signal (idler) photons from the entangled pairs are directed to Alice (Bob), where corresponding projection measurements are conducted on them. A set of optical components, including QWP, HWP, and QWP, is employed to construct the Wave Plate Group (WPG), facilitating the creation of any projection measurement in a two-dimensional subspace. This approach is chosen because any high-dimensional projection measurement can be decomposed into several two-dimensional subspace projection measurements. The measured signal (idler) photons are collected into coupler C1 (C2), and coincidence counts are determined by using a time-to-digital converter (ID800). PBS: Polarizing beam splitter. D-PBS: Dichroic polarizing beam splitter. BD: Beam displacer. DM: Dichroic mirror. D-HWP: Dichroic Half wave plate. HWP: Half wave plate. QWP: Quarter wave plate. WPG: Wave plate group. BS: Beam splitter. PPKTP: Periodically poled potassium titanium phosphate.}
		\label{fig:setup}
		\vspace{-0.2cm}
	\end{figure*}
	
	\textit{Randomness certification in MIMO scenario. ---}Consider a scenario where two parties, Alice and Bob, are located in distant laboratories and share a quantum state $\rho_{AB}$. Both of their measurement devices are untrusted, and each of them performs one out of $m$ measurements that yields $n$ possible outcomes. 
	The measurements and outcomes are labeled as $x,y\in\{1,\cdots,m\}$ and $a,b\in\{0,\cdots,n-1\}$ respectively, and hence the corresponding elements of positive operator valued measure (POVM) measurement are labeled as $M_{a|x}$ ($M_{b|y}$) for Alice (Bob). 
	The correlations shared between them are characterized by the joint probabilities $P(A_x=a,B_y=b) = {\rm Tr} [M_{a|x}\otimes M_{b|y} \rho_{AB}]$, and give a value $I^{\rm obs}$ for a given Bell expression $I = \sum_{abxy}I_{abxy}P(A_x = a, B_y = b)$. A local behavior should satisfy the corresponding Bell inequality $I\leq \mathcal{C}$. Here $\mathcal{C} = \max_{\vec P \in \mathcal{L}} I$ is the classical bound, and the set $\mathcal{L}$ is a polytope, where the extreme points obey the local deterministic behaviors~\cite{rmp_2014_nonlocality}. Therefore, the Bell inequality can be used to test the nonlocal behaviors between Alice and Bob. 
	Within this scenario, an eavesdropper Eve optimizes her strategy to give a guess $(e,e')$ for the outcomes of certain target measurements $(x^*,y^*)$, while being constrained by $I^{\text{obs}}$ and quantum theory. 
	Hence, the guessing probability is defined by the subsequent optimization problem~\cite{science_2018_jianweiwang,nature_2010_random}:
	\begin{equation}
	\begin{aligned}
	\label{eq:definition}
	P_g(x^*,y^*) &\coloneqq \max_{\left\{P(A_{x}=a,B_{y} = b,e,e')\right\}} \sum_{e,e'} P(A_{x^*} = e,B_{y^*} = e',e,e') \\
	\text{s.t.}  ~~&\sum_{a,b,e,e',x,y}I_{abxy}P(A_x = a,B_y = b,e,e') = I^{\rm obs},\\
	& ~~~P(A_x = a,B_y = b,e,e') \in Q \quad \forall e,e',
	\end{aligned}
	\end{equation}
	where the optimization variables are the collection of probability distributions $P(A_x = a, B_y = b,e,e')$ for Alice, Bob, and Eve. Here $Q$ represents the set of quantum correlations that is not a polytope, making it difficult to be characterized~\cite{rmp_2014_nonlocality}. 
	However, one can relax the optimization problem Eq.~\eqref{eq:definition} by changing $Q$ by a superset $Q_k\supset Q$, such as the Navascu\'{e}s, Pironio and Ac\'{i}n (NPA) hierarchy~\cite{prl_2007_npa,njp_2008_npa}, which can efficiently be characterized via semidefinite programming techniques. The parameter $k$ is an integer that trades off computational efficiency for accuracy.
	By solving this problem, some lower bounds of the certifiable randomness are obtained by the min-entropy $H_{\min} \coloneqq -\log_2(P_g)$~\cite{Ieee_2009_Hmin}, where $P_g = \min_{x,y}P_g(x,y)$ since we are discussing the minimal quantum resource required for randomness certification. We can also give its upper bounds to precisely determine the certifiable randomness by the see-saw algorithm~\cite{prl_2024_yi,supp}.
	\nocite{cvx1,cvx2,qetlab,mosek,prr_2021_quantifyIC}
	
	\textit{Nonlocality vs. Randomness in MIMO Scenario. ---} To certify nonzero randomness based on Eq.~\eqref{eq:definition}, the given Bell inequality must be violated, otherwise Eve can always find a deterministic strategy to give unit guessing probability~\cite{prl_1982_fine,prl_2012_acin}.
	We prove that, for any Bell inequality with two inputs, the violation is \textit{necessary and sufficient} for certifying nonzero randomness. The sufficiency, however, does not generally hold with more inputs.
	
	The idea is the following: if no randomness can be certified from the outcomes of target measurements $x^*$ and $y^*$, then these outcomes are always equal to Eve's guesses, thereby the observed correlation can be explained by a scenario where the measurements $M_{ee'=ab}$ determined by $x^*$ and $y^*$ are performed by Eve. Since Eve's measurement is locally separated with Alice's and Bob's, the measurements implemented by the untrusted parties can be explained by a certain structure, where the target measurement is compatible with every other measurement respectively, e.g., the measurement $x^*$ is compatible with other measurements $x\neq x^*$. Therefore, in the two-input scenario, the inability to certify randomness indicates that the set of Alice's measurements is compatible, and likewise for Bob, resulting in no Bell violation that can be observed~\cite{supp}.
	
	However, due to the compatibility structures becomes complex with the increase of measurements, the sufficiency is generally not applicable to Bell inequalities with more than two inputs, i.e., there are instances where a MIMO Bell inequality is violated yet no randomness can be assured. But for some certain inequalities whose violation can exclude the above unique structure of measurement incompatibility~\cite{supp}, for example, SATWAP inequalities, the equivalence between nonlocality and certifiable randomness still holds with more than two inputs.
	
	In the following, we examine two typical classes of MIMO Bell inequalities, the facet inequalities $I_{mn}^{\rm F}$ and the SATWAP inequalities $I_{mn}^{\rm S}$ with different $m$ inputs and $n$ outputs, which are commonly tested in high-dimensional systems.
	We observe that different from the performance of SATWAP inequalities, the violation of the facet inequality $I_{42}^{\rm F}$ is not sufficient for certifying randomness. We choose the SATWAP expressions~\cite{prl_2017_SATWAP} with $(m, n)=(2,2)$, $(3,2)$, $(2,4)$, $(3,4)$:
	\begin{equation}
	\begin{aligned}
	\label{eq:satwap}
	I_{mn}^{\rm S}&=\sum_{k=0}^{\lfloor n / 2\rfloor-1}\left(\alpha_k \mathbb{P}_k-\beta_k \mathbb{Q}_k\right),
	\end{aligned}
	\end{equation}
	where $\mathbb{P}_k=\sum_{i=1}^m\left[P\left(A_i=B_i+k\right)+P\left(B_i=A_{i+1}+k\right)\right]$, $\mathbb{Q}_k =\sum_{i=1}^m\left[P\left(A_i=B_i-k-1\right)+P\left(B_i=A_{i+1}-k-1\right)\right]$ with $A_{m+1}\equiv A_1+1$. Here the probability $P(A_a = B_b + k)$ means that the outcomes of measurement $A_a$ and $B_b$ differ modulo $d$ by $k$. The parameters $\alpha_k$, $\beta_k$ and its classical bound as well as its quantum bound can be found in the Supplemental Material~\cite{supp}. 
	For the facet inequalities, we select the CHSH inequality $I_{22}^{\rm F}$ as well as the Collins-Gisin-Linden-Massar-Popescu inequality~\cite{prl_2002_CGLMP} $I_{24}^{\rm F}$ with $(m,n)=(2,4)$. The latter actually shares the same expression as Eq.~\eqref{eq:satwap} but with $\alpha_k = \beta_k = [1-2k/3]$. Further, we choose the $I_{4422}^4$ inequality~\cite{prl_2010_belltest} for $I_{42}^{\rm F}$:
	\begin{equation}
	\begin{aligned}
	I_{42}^{\rm F} & = I_{\mathrm{CH}}^{(1,2 ; 1,2)}+I_{\mathrm{CH}}^{(3,4 ; 3,4)}-I_{\mathrm{CH}}^{(2,1 ; 4,3)}-I_{\mathrm{CH}}^{(4,3 ; 2,1)} - P\left(A_2 = 1 \right)\\ 
	& ~~~ -P\left(A_4 = 1\right)-P\left(B_2 = 1\right)-P\left(B_4 = 1\right)\leq 0,
	\end{aligned}
	\end{equation}
	where $I_{\mathrm{CH}}^{(i, j ; p, q)} = P(A_i = B_p = 1)+P(A_j = B_p = 1)+P(A_i = B_q = 1) - P(A_j = B_q = 1)-P(A_i = 1)-P(B_p = 1)$. This inequality is recognized for its ability to tolerate a detection efficiency as low as $61.8\%$~\cite{prl_2010_belltest,prl_2022_belltest}, in contrast to the optimal threshold of detection efficiency ($2/3$) in the two-input and two-output scenario.
	
	To compare their different behaviors towards randomness, we normalize the expressions $I_{mn}^{\rm F}$ and $I_{mn}^{\rm S}$ with the same classical bound of $0$ as well as the quantum bound of $1$. In general, the value of a Bell expression that is below the quantum bound can be achieved by performing POVMs on a quantum state without restricting the dimension of systems~\cite{rmp_2014_nonlocality}. Instead of using POVMs, in our experiment we show that the expression values can exceed the classical bound in a four-dimensional system by designing the schemes consisting of pure states and projective measurements.
	Then we substitute these values into Eq.~\eqref{eq:definition} to certify the randomness within the current system, thus exploring the connection between Bell nonlocality tested by various inequalities and the level of extractable randomness. 
	
	\begin{figure*}[ht!]
		\centering
		\includegraphics[width=0.83\textwidth]{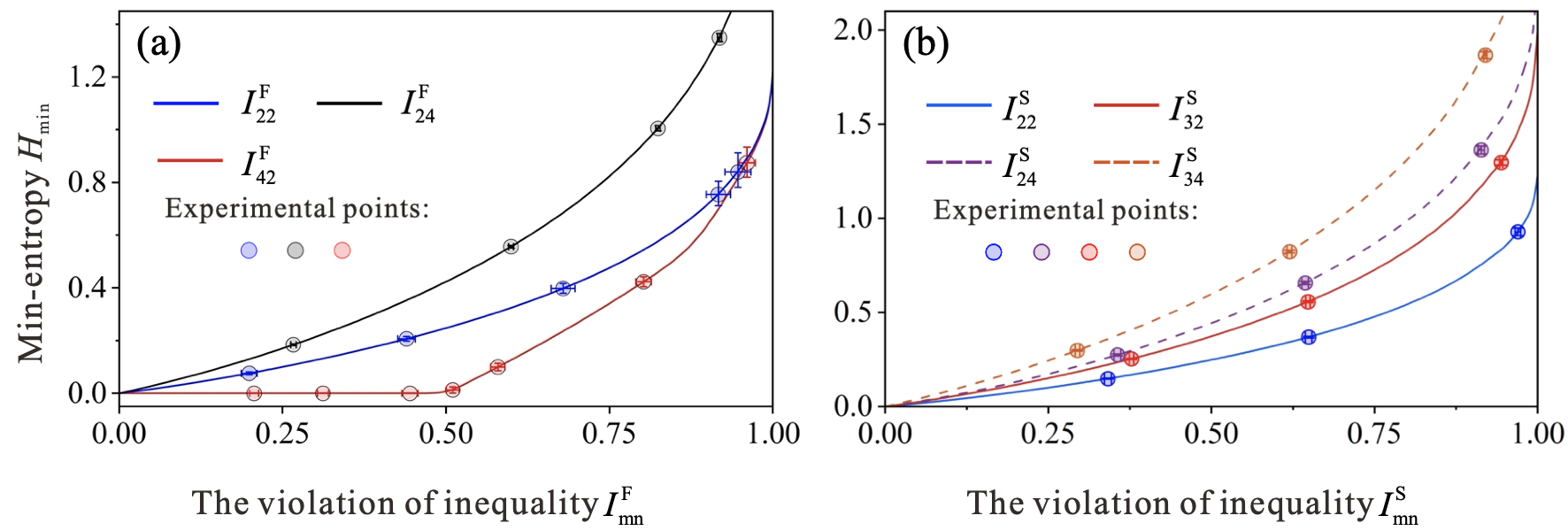}
		\caption{\textmd{The horizontal axis represents the violation of inequality $I_{mn}^{\text{F}}$ (or $I_{mn}^{\text{S}}$) (normalized), and the vertical axis represents the lower bounds of certifiable randomness that the system can certify under the premise of testing Bell nonlocality.} Figures (\textbf{a}) and (\textbf{b}) respectively show the violations of $I_{22}^{\text{F}}$, $I_{24}^{\text{F}}$, and $I_{42}^{\text{F}}$ of the facet inequality $I_{mn}^{\text{F}}$, and $I_{22}^{\text{S}}$, $I_{32}^{\text{S}}$, $I_{24}^{\text{S}}$ and $I_{34}^{\text{S}}$ of the SATWAP inequality $I_{mn}^{\text{S}}$ obtained from experiments. For all the inequalities except the $I_{42}^{\text{F}}$ inequality, the lower bounds of the certifiable randomness were numerically determined at the level $1+AB$ of the NPA hierarchy. For the $I_{42}^{\text{F}}$ inequality, the certifiable randomness is bounded at the level of $1+AB+AAB$~\cite{supp}. We note that its upper bound and zero coincide up to numerical precision ($10^{-9}$) when $I_{42}^{\text{F}}\in [0,0.499]$. The error of the Bell value is calculated by simulating photons with 1000 Poisson distribution.  The error of randomness corresponds to $-(+)1\sigma$ confidence interval for the Bell value.}
		\label{fig:exp result}
	\end{figure*}
	
	\textit{Experimental Setup. ---} Our experimental platform can generate various two-dimensional and four-dimensional entangled states and perform arbitrary projection measurements. The continuous-wave laser with a wavelength of 775~nm is split into two parallel paths (`a', `b') with 4~mm separation \cite{PhysRevLett.129.030502}. The PPKTP crystal is pumped in a Sagnac structure, undergoing the type-II spontaneous parametric down-conversion (SPDC) process to generate correlated (signal and idler) photon pairs. These photons are encoded with both path and polarization degrees of freedom (DOFs) \cite{PhysRevLett.125.090503}, i.e. $|a_H\rangle$, $|a_V\rangle$, $|b_H\rangle$ and $|b_V\rangle$ are encoded as $|0\rangle$, $|1\rangle$, $|2\rangle$ and $|3\rangle$,  respectively, where the symbol `$H$' (or `$V$') denotes horizontally (or vertically) polarized photons on the path. The resulting four-dimensional entangled states are expressed as 
	\begin{equation}
	\begin{aligned}
	|\Psi\rangle=&\cos (\theta_1)\cos (\theta_2)|00\rangle+\cos(\theta_1)\sin (\theta_2) |11\rangle+\\
	&\sin (\theta_1)\cos (\theta_3)|22\rangle+\sin (\theta_1)\sin (\theta_3)|33\rangle,
	\end{aligned}
	\end{equation}
	where parameters $\theta_1$, $\theta_2$, and $\theta_3$ can be respectively adjusted by HWP1-HWP3 to prepare the desired entangled states. Notably, by setting $\theta_1=0$ or $\theta_3=0$, some two- or three-dimensional entangled states can also be obtained by designing the remaining parameters.
	
	The signal (idler) photons are distributed to Alice (Bob) to perform local measurements. Performing high-dimensional projection measurements involves more complex operations and additional optical components compared to two-dimensional projection measurements \cite{PhysRevLett.127.110505,PhysRevLett.73.58}. As shown in Fig. \ref{fig:setup}, each WPG is composed of QWP, HWP, and QWP. Taking Alice as an example, WPG1 and WPG2, combined with BD2 are utilized to perform the projection measurement within the two-dimensional subspace extended by ($\{|0\rangle, |1\rangle\}$) and ($\{|2\rangle, |3\rangle\}$), respectively. Subsequently, using WPG3 and PBS1, four-dimensional projection measurements were completed. Similarly, by adjusting the angle of WPGs, Alice can also perform any two- or three-dimensional projection measurement. Following the removal of noise through a bandpass filter ($1550\pm15$nm) and a long pass filter (1200~nm), the measured photons are collected in coupler C1 (C2) and detected by the superconducting detector.
	
	The entangled source generates approximately 5,000 pairs of entangled photons per second, with a total of 30 records per setting. The details regarding the target entangled states and the corresponding projection measurements can be found in the Supplemental Material~\cite{supp}.
	
	\textit{Results. ---} Firstly, we measure the facet Bell inequalities with two inputs and two outputs $I_{22}^\text{F}$, two inputs and four outputs $I_{24}^\text{F}$, and four inputs and two outputs $I_{42}^\text{F}$, respectively. 
	As shown in Fig.~\ref{fig:exp result} (a), when the number of measurements is $m=2$, that is, for inequalities $I_{22}^\text{F}$ and $I_{24}^\text{F}$, both Bell nonlocality ($I_{mn}^\text{F}>0$) and randomness ($H_{\min}>0$) in the experimental system occur simultaneously, which aligns with the numerical results from various previous works~\cite{nature_2010_random,quantum_2018_multipartite,pra_2013_multipartite,njp_2014_the,jpa_2014_the}. However, when the number of measurements basis increases to $m=4$, that is, for inequality $I_{42}^\text{F}$, certifying randomness from the experimental system is not feasible unless the violation of $I_{42}^\text{F}$ is larger than 0.499. This indicates that the Bell nonlocality tested by the facet inequalities $I_{mn}^\text{F}$ is not always a sufficient condition for certifying the randomness; rather, it depends on the selected number of measurements. The accurate values of all $I_{mn}^{\text{F}}$ class inequalities obtained and the corresponding certified randomness are listed in the Supplemental Material~\cite{supp}.
	
	Next, we measure the SATWAP inequalities $I_{mn}^\text{S}$ and select the number of measurements $m=2, 3$ and outputs $n=2, 4$, respectively. The violations (normalized) of inequality $I_{mn}^\text{S}$ obtained in the experiment and the corresponding certified randomness are presented in Fig.~\ref{fig:exp result} (b), whose accurate values are listed in the Supplemental Material~\cite{supp}. Clearly, regardless of the number of inputs $m$, Bell nonlocality ($I_{mn}^\text{S}>0$) and randomness ($H_{\min}>0$) always occur simultaneously. 
	
	The fact that the SATWAP inequality is sensitive for certifying randomness can be analyzed by testing the measurement incompatibility~\cite{prl_2019_detectICstructure} on the untrusted parties. We prove that the measurements $x^*$ and $y^*$ are at least incompatible with one of the rest measurements when the SATWAP inequalities are violated, but not to the case of facet inequalities~\cite{supp}. In addition, we show that higher randomness can be certified in the case of $n = 4$, where we perform projective measurements with four outputs to a four-dimensional system. The private randomness with a generation rate of $1.867\pm0.018$ bits per every photon pair is obtained in the scenario of $I_{34}^\text{S}$. Further, since the quantum bounds of SATWAP inequalities can be reached by performing specific measurements on the maximally $d$-dimensional entangled state, we analytically provide upper bounds on the certifiable randomness when these inequalities are maximally violated~\cite{supp}.
	
	\textit{Discussion and Conclusion. ---} Generating DI random numbers requires loophole-free Bell testing, which will pose significant challenges to the experiment. Moreover, finite-size experimental data can also influence the generation rate of random numbers, and it is necessary to consider smooth minimum entropy~\cite{arnon2019simple} and use a corresponding Toeplitz matrix hashing extractor to effectively extract random numbers~\cite{nature_2018_ustc}. Specifically, to avoid the assumption of fair sampling, the detection efficiency of entangled photon pairs must be higher than a certain threshold. This is a huge challenge in the experiment, especially in the photon Bell loophole-free experiment, where fiber loss and detector efficiency pose great difficulties\cite{prl_2022_belltest,PhysRevLett.115.250401}. As is well known, increasing the number of inputs can significantly reduce the threshold of detection efficiency required to violate Bell inequalities. For example, the facet inequality $I_{42}^{\text{F}}$ is notable for its ability to tolerate a detection efficiency lower down to $61.8\%$~\cite{prl_2010_belltest,prl_2022_belltest}. It then motivates us to compare the abilities of different inequalities to resist detection loss~\cite{prl_2013_belltest} when certifying randomness. We find that, even though the inequality $I_{42}^{\text{F}}$ exhibits a lower detection efficiency tolerance compared to the SATWAP inequality $I_{32}^{\text{S}}$ in expressing Bell nonlocality, this advantage, however, reverses when it comes to randomness certification. Explicitly, the violation of $I_{42}^{\text{F}}$ cannot exceed $0.499$ when the detection efficiency falls below $90.6\%$, thereby rendering no randomness certifiable from Eq.~\eqref{eq:definition}. In contrast, when the detection efficiency is larger than $80.0\%$, $I_{32}^\text{S}$ can be violated to demonstrate Bell nonlocality as well as the certifiable randomness~\cite{supp}, as the SATWAP inequality is more sensitive for this task. In this context, one can certify randomness based on the SATWAP inequality within a broader region of detection efficiency.
	
	In summary, we deepen the understanding of the connection between nonlocality and quantum randomness, which differs from the previous results that the randomness is certified once the nonlocal behavior is detected.
	We have proven that not all Bell nonlocal states can be used to certify randomness; the equivalence only holds for two-input Bell inequalities, but not for multiple inputs. Our work greatly facilitates the development of high-performance device-independent quantum information tasks~\cite{prl_2022_belltest,PhysRevX.12.041023}.

	\begin{acknowledgments}
		This work was supported by the National Natural Science Foundation of China (No.~12374338, No.~12174367, No.~12204458,  No.~11904357, No.~12125402, No.~12350006, and No.~62322513), the Innovation Program for Quantum Science and Technology (No. 2021ZD0301200 and No. 2021ZD0301500), and Beijing Natural Science Foundation (Grant No. Z240007). This work was partially carried out at the USTC Center for Micro and Nanoscale Research and Fabrication.
		J.T. acknowledges the support received by the Dutch National Growth Fund (NGF), as part of the Quantum Delta NL programme, as well as the support received from the European Union’s Horizon Europe research and innovation programme through the ERC StG FINE-TEA-SQUAD (Grant No. 101040729). The views and opinions expressed here are solely those of the authors and do not necessarily reflect those of the funding institutions. Neither of the funding institution can be held responsible for them.
	\end{acknowledgments}
	
\bibliography{Ref_PRL_round1}
\end{document}


\title{Supplemental Material: Randomness versus Nonlocality in Multi-input and Multi-output Quantum Scenario}

\author{Chao Zhang}
\thanks{These authors contributed equally to this work.}
\affiliation
{CAS Key Laboratory of Quantum Information, University of Science and Technology of China, Hefei 230026, China}
\affiliation{CAS Synergetic Innovation Center of Quantum Information $\&$ Quantum Physics, University of Science and Technology of China, Hefei 230026, China}
\author{Yi Li}
\thanks{These authors contributed equally to this work.}
\affiliation{State Key Laboratory for Mesoscopic Physics, School of Physics, Frontiers Science Center for Nano-optoelectronics, $\&$ Collaborative Innovation Center of Quantum Matter, Peking University, Beijing 100871, China}
\affiliation{Beijing Academy of Quantum Information Sciences, Beijing 100193, China}
\author{Xiao-Min Hu}
\thanks{These authors contributed equally to this work.}
\affiliation
{CAS Key Laboratory of Quantum Information, University of Science and Technology of China, Hefei 230026, China}
\affiliation{CAS Synergetic Innovation Center of Quantum Information $\&$ Quantum Physics, University of Science and Technology of China, Hefei 230026, China}
\author{Yu Xiang}
\email{xiangy.phy@pku.edu.cn}
\affiliation{State Key Laboratory for Mesoscopic Physics, School of Physics, Frontiers Science Center for Nano-optoelectronics, $\&$ Collaborative Innovation Center of Quantum Matter, Peking University, Beijing 100871, China}
\affiliation{Collaborative Innovation Center of Extreme Optics, Shanxi University, Taiyuan, Shanxi 030006, China}
\author{Chuan-Feng Li}
\affiliation
{CAS Key Laboratory of Quantum Information, University of Science and Technology of China, Hefei 230026, China}
\affiliation{CAS Synergetic Innovation Center of Quantum Information $\&$ Quantum Physics, University of Science and Technology of China, Hefei 230026, China}
\author{Guang-Can Guo}
\affiliation
{CAS Key Laboratory of Quantum Information, University of Science and Technology of China, Hefei 230026, China}
\affiliation{CAS Synergetic Innovation Center of Quantum Information $\&$ Quantum Physics, University of Science and Technology of China, Hefei 230026, China}
\author{Jordi Tura}
\affiliation{$\langle aQa^L\rangle$ Applied Quantum Algorithms Leiden, The Netherlands}
\affiliation{Instituut-Lorentz, Universiteit Leiden, P.O. Box 9506, 2300 RA Leiden, The Netherlands}
\author{Qihuang Gong}
\affiliation{State Key Laboratory for Mesoscopic Physics, School of Physics, Frontiers Science Center for Nano-optoelectronics, $\&$ Collaborative Innovation Center of Quantum Matter, Peking University, Beijing 100871, China}
\affiliation{Collaborative Innovation Center of Extreme Optics, Shanxi University, Taiyuan, Shanxi 030006, China}
\affiliation{Peking University Yangtze Delta Institute of Optoelectronics, Nantong, Jiangsu 226010, China}	
\affiliation{Hefei National Laboratory, Hefei 230088, China}
\author{Qiongyi He}
\email{qiongyihe@pku.edu.cn}
\affiliation{State Key Laboratory for Mesoscopic Physics, School of Physics, Frontiers Science Center for Nano-optoelectronics, $\&$ Collaborative Innovation Center of Quantum Matter, Peking University, Beijing 100871, China}
\affiliation{Collaborative Innovation Center of Extreme Optics, Shanxi University, Taiyuan, Shanxi 030006, China}
\affiliation{Peking University Yangtze Delta Institute of Optoelectronics, Nantong, Jiangsu 226010, China}	
\affiliation{Hefei National Laboratory, Hefei 230088, China}
\author{Bi-Heng Liu}
\email{bhliu@ustc.edu.cn}
\affiliation{CAS Key Laboratory of Quantum Information, University of Science and Technology of China, Hefei 230026, China}
\affiliation{CAS Synergetic Innovation Center of Quantum Information $\&$ Quantum Physics, University of Science and Technology of China, Hefei 230026, China}

\maketitle

\section{Nonlocality as resources for certifying randomness}
In this section, we first discuss the connection between nonlocality and certifiable randomness by linking the structure of measurement incompatibility with the Local-Hidden Variable (LHV) model. Secondly, to accurately certify randomness, we use the see-saw algorithm to calculate the upper bound of certifiable randomness. Then we test the structure of measurement incompatibility based on the SATWAP inequalities, showing that the chosen SATWAP inequalities are necessary and sufficient for certifying randomness. Finally, we calculate the threshold of detection efficiency for the SATWAP inequality with three inputs and two outputs, and compare it with the $I_{4422}^4$ inequality in the context of randomness certification. We denote $P(A_x = a, B_y = b)$ as $P(ab|xy)$ in this section and all numerical SDP calculations were performed using the CVX package for MATLAB~\cite{cvx1,cvx2}, along with the library QETLAB~\cite{qetlab} as well as the MOSEK solver~\cite{mosek}.

\subsection{Nonlocality versus certifiable randomness}

\subsubsection{A violation of Bell inequality is necessary for certifying randomness}
When Alice and Bob observe a value $I^{\rm obs}$ of a given Bell expression $I = \sum_{abxy}I_{abxy}P(ab|xy)$, and it does not exceed the classical bound $\mathcal{C} = \max_{\vec P \in \mathcal{L}} I$. Then this value can be achieved by a distribution $P^{\rm obs}(ab|xy)$, which can be written as the LHV model~\cite{rmp_2014_nonlocality}:
\begin{equation}
    P^{\rm obs}(ab|xy) = \sum_{\lambda}p_{\lambda} D_{\rm Local} (ab|xy,\lambda).
\end{equation}
Here each $\lambda$ corresponds to two strings of outcomes $\lambda = (a_{x=1},a_{x=2},\cdots,a_{x=m}; b_{y=1},b_{y=2},\cdots,b_{y=m})$, and the deterministic behavior is
\begin{equation}
    \begin{aligned}
        D_{Local}(ab|xy,\lambda) = \begin{cases}1, & \text { if } a = a_x \text { and } b = b_y,\\
        0,&\text{ otherwise.}\end{cases}
    \end{aligned}
\end{equation}
Then we reconstruct the distribution as
\begin{equation}
    \begin{aligned}
        P^{\rm obs}(ab|xy) = \sum_{\lambda}p_{\lambda} D_{\rm Local} (ab|xy,\lambda) = \sum_{e,e'}\sum_{\lambda\in\Lambda^{(ee')}}p_{\lambda} D_{\rm Local} (ab|xy,\lambda),
    \end{aligned}
\end{equation}
where $\Lambda^{(e,e')}:= \{ \lambda|\lambda{(x^*;y^*)} = (e;e')\}$, $\forall e,e'$. Here we denote that $\lambda(\cdot;\cdot)$ is a function, which takes two inputs and gives two outputs. Each input takes from the set $\{1,\cdots,m\}$, and produces output within the range of $\{0,\cdots,n-1\}$.
Hence, based on the observed distribution, one cannot rule out a situation in which Eve gives unit guessing probability. For instance, Alice, Bob, and Eve share a tripartite state $\rho_{ABE} = \sum_{e,e'} |e,e'\rangle_E\langle e,e'| \otimes  \sum_{\lambda \in \Lambda^{(e,e')}} p_\lambda \rho_{AB}^{\lambda}$, and perform measurements $\{M_{a|x}\}_{a,x}$, $\{M_{b|y}\}_{b,y}$ and $\{M_{e,e'}\}_{e,e'}$ respectively, where
\begin{equation}
    \begin{aligned}
        \rho_{AB}^{\lambda} &= |a_1\rangle\langle a_1| \otimes \cdots \otimes |a_{m}\rangle\langle a_{m}| \otimes |b_1\rangle\langle b_1|\otimes \cdots\otimes |b_{m}\rangle\langle b_{m}|,\\
        M_{a|x} &= \mathbb{I}_1\otimes \cdots\otimes |a\rangle_x\langle a| \otimes \cdots \otimes \mathbb{I}_{m}, \quad \forall a,x, \quad \sum_{a}|a\rangle\langle a| = \mathbb{I},\\
        M_{b|y} &= \mathbb{I}_1\otimes \cdots\otimes |b\rangle_y\langle b| \otimes \cdots \otimes \mathbb{I}_{m}, \quad \forall b,y, \quad \sum_{b}|b\rangle\langle b| = \mathbb{I},\\
          M_{e,e'} &= |e,e'\rangle\langle e,e'|, \quad \sum_{e,e'}|e,e'\rangle\langle e,e'| = \mathbb{I}.
    \end{aligned}
\end{equation}
In this case, the states $\rho^\lambda_{AB}$ and these measurements yield the deterministic behavior for each $\lambda$. One can easily check that this example is compatible with distribution $P^{\rm obs}(ab|xy)$, as well as with the observed value $I^{\rm obs}$. Hence, while Eve's strategy is compatible with the observed value $I^{\rm obs}$, her guesses are always equal to Alice's and Bob's outcomes. It means one cannot certify randomness from Eq.~(1) in the manuscript when the given Bell inequality is not violated. We note that one can certify more randomness without the need to process the observed data into an inequality, i.e., the randomness is directly quantified by the observed distribution~\cite{njp_2014_the}. But in this work, we certify randomness just through the amount of violation of the given Bell inequality, like the works~\cite{science_2018_jianweiwang,nature_2010_random,quantum_2018_multipartite}.
\subsubsection{A violation of two-input Bell inequality is sufficient for certifying randomness}
In the case of $m = 2$, we denote that $x\in\{x^*,\bar{x}^*\}$ and $y\in\{y^*,\bar{y}^*\}$. Then $P_g(x^*,y^*) = 1$ in Eq.~(1) in the main text means 
\begin{equation}
    \begin{aligned}
        P_g(x^*,y^*) = \sum_{e,e'} p(e,e',e,e'|x^*,y^*) = \sum_{e,e'} p(e,e'|x^*,y^*,e,e')p(e,e') = 1,
    \end{aligned}
\end{equation}
where $p(a,b,e,e'|x,y) = p(a,b|x,y,e,e')p(e,e'|x,y) = p(a,b|x,y,e,e')p(e,e')$. Thus, we have $p(a,b|x^*,y^*,e,e') = \delta_{a,e}\delta_{b,e'}$, $\forall e,e',a,b$. Based on the no-signaling condition,
\begin{equation}
    \begin{aligned}
        \sum_{a}p(a,b|x^*,y^*,e,e') = \sum_{a}\delta_{a,e}\delta_{b,e'} = \delta_{b,e'} = \sum_{a}p(a,b|\bar{x}^*,y^*,e,e'), \qquad \forall b,\bar{x}^*,y^*,e,e',\\
        \sum_{b}p(a,b|x^*,y^*,e,e') = \sum_{b}\delta_{a,e}\delta_{b,e'} = \delta_{a,e}  = \sum_{b}p(a,b|x^*,\bar{y}^*,e,e'), \qquad \forall b,\bar{y}^*,x^*,e,e',
    \end{aligned}
\end{equation}
we have $p(a,b| x^*, \bar y^*,e,e') = \delta_{a,e}p(a,b|x^*, \bar y^*,e,e')$ and $p(a,b|\bar x^*, y^*,e,e') = \delta_{b,e'}p(a,b|\bar x^*,y^*,e,e')$.

Since Eve only performs one POVM, the state of Alice, Bob and Eve can be written as classical-quantum state $\rho_{EAB} = \sum_{e,e'}p(e,e')|e\rangle_{E_1}\langle e|\otimes |e'\rangle_{E_2}\langle e'|\otimes \rho_{AB}^{ee'}$ and  Eve can be regarded as two local parts ($E_1$ and $E_2$) without loss of generality. Therefore, no certifiable randomness means
\begin{equation}
	\begin{aligned}
		{\rm Tr}[M_{e}^1\otimes M_{e'}^2\otimes M_{a|x} \otimes M_{b|y}\rho_{E_1E_2AB}] = \begin{cases} \delta_{e,a}\delta_{e',b} p(a,b,e,e'|x,y), & \text { for } x=x^* \text { and } y=y^*, \\ 
		\delta_{e,a}p(a,b,e,e'|x,y), &\text { for } x=x^* \text { and } y=\bar{y}^*,\\ 
		\delta_{e',b}p(a,b,e,e'|x,y), &\text { for } x=\bar{x}^* \text { and } y={y}^*,\\ 
		p(a,b,e,e'|x,y), & \text { otherwise, }\end{cases} \qquad \forall e,e',a,b,
	\end{aligned}
\end{equation}
which means
\begin{equation}
    \begin{aligned}
        \sum_{k,l} \text{Tr} \left[ M_{a}^1 \otimes M_{b}^2\otimes M_{l| x^*}\otimes M_{k| y^*} \rho_{E_1E_2AB}\right]  &= \sum_{k,l} \text{Tr} \left[ M_{l}^1 \otimes M_{k}^2\otimes M_{a|x^*} \otimes M_{b|y^*} \rho_{E_1E_2AB} \right], \quad \forall a,b\\
        \sum_{l} \text{Tr} \left[ M_{a}^1 \otimes M_{e'}^2\otimes M_{l| x^*}\otimes M_{b|\bar y^*} \rho_{E_1E_2AB}\right] &= \sum_{l} \text{Tr} \left[ M_{l}^1 \otimes M_{e'}^2\otimes M_{a|x^*} \otimes M_{b|\bar y^*} \rho_{E_1E_2AB}\right], \quad \forall a,b,e',\\\sum_{k} \text{Tr} \left[ M_{e}^1 \otimes M_{b}^2\otimes M_{a|\bar x^*}\otimes M_{k|y^*} \rho_{E_1E_2AB} \right] &= \sum_{k} \text{Tr} \left[ M_{e}^1 \otimes M_{k}^2\otimes M_{a|\bar x^*} \otimes M_{b|y^*}\rho_{E_1E_2AB} \right], \quad \forall a,b,e\\ 
    \end{aligned}
\end{equation}
Then from the observed joint probability distribution, which can be written as
\begin{equation}
    \begin{aligned}
        P(a,b|\bar x^*,y^*) &= \sum_{k,e} \text{Tr} \left[ M_{e}^1 \otimes M_{b}^2\otimes M_{a|\bar x^*}\otimes M_{k|y^*} \rho_{E_1E_2AB}\right] = \text{Tr} \left[ I_{E_1}\otimes M_{b}^2\otimes M_{a|\bar x^*}\otimes I_B \rho_{E_1E_2AB}\right], \quad \forall a,b,\\
        P(a,b|x^*,\bar y^*) &= \sum_{l,e'} \text{Tr} \left[ M_{a}^1 \otimes M_{e'}^2\otimes M_{l|x^*}\otimes M_{b|\bar y^*} \rho_{E_1E_2AB}\right] = \text{Tr} \left[M_{a}^1\otimes I_{E_2}\otimes I_A\otimes M_{b|\bar{y}^*} \rho_{E_1E_2AB}\right], \quad \forall a,b,\\ 
        P(a,b|x^*,y^*) &= \sum_{k,l} \text{Tr} \left[ M_{a}^1 \otimes M_{b}^2\otimes M_{l| x^*}\otimes M_{k| y^*} \rho_{E_1E_2AB}\right]   = \text{Tr} \left[ M_{a}^1 \otimes M_{b}^2\otimes I_A \otimes I_B \rho_{E_1E_2AB} \right], \quad \forall a,b,\\
        P(a,b|\bar x^*,\bar y^*) &= \text{Tr} \left[ I_{E_1} \otimes I_{E_2}\otimes M_{a|\bar x^*} \otimes M_{b|\bar y^*} \rho_{E_1E_2AB} \right], \quad \forall a,b,
    \end{aligned}
\end{equation}
Therefore, in the two inputs scenario, Alice's measurements ($\{M_a^1\otimes I_A , I_{E_1}\otimes M_{a|\bar x^*}\}$) as well as Bob's measurements ($\{M_b^2\otimes I_B , I_{E_2}\otimes M_{b|\bar y^*}\}$) are compatible if there is no certifiable randomness. Both Alice's and Bob's measurements are compatible, which means the observed joint probability distribution cannot be nonlocal. Hence no two-input Bell inequality can be violated by this local behavior. Nonetheless, in the case of $m\geq 3$, the additional measurement could express nonlocality with the measurements other than $x^*$ or $y^*$ but are not involved in generating randomness. This proof can be easily generalized to multipartite cases.

\subsection{Upper bounds of certifiable randomness}
The upper bound of the min-entropy can be calculated by fixing the dimension of each subsystem and solving the following problem:
\begin{equation}
    \begin{aligned}
    \label{suppeq:upperbound}
			\max \quad &\sum_{e,e^\prime}\operatorname{Tr}\left[  \left(  M_{a = e|x^*} \otimes M_{b = e^\prime|y^*} \right)  \sigma_{ AB}^{ee'}  \right]\\ 
   \text{w.r.t.} \quad & \left\{M_{a|x}\right\}_{a,x}, \left\{M_{b|y}\right\}_{b,y},\{\sigma_{AB}^{ee'}\}_{e,e'}\\ 
			\operatorname{ s.t.} \quad& \sum_{a,b,x,y,e,e^\prime} I_{abxy}\operatorname{Tr}\left[\left( M_{a|x}\otimes M_{b|y} \right)\sigma_{AB}^{ee^\prime }\right] = I^{\rm obs}, \quad \forall a,b,x,y,\\
			& \left\{M_{a|x}\right\}_a, \left\{M_{b|y}\right\}_b\in {\rm POVM}, \quad \forall x,y, \\
   & \sigma_{AB}^{ee'}\geq 0,\quad \forall e,e', \quad \sum_{ee'} {\rm Tr}\left(\sigma_{AB}^{ee'}\right) = 1.
		\end{aligned}
\end{equation}
Here we use $d_A,d_B$ to denote the dimensions of Alice and Bob respectively. However, even by fixing the dimensions, the problem Eq.~\eqref{suppeq:upperbound} is neither an SDP problem nor a linear problem. Here we use the see-saw algorithm to convert this problem into three SDPs. 

Firstly, we set the POVMs $\{ M_{a|x} ^{(0)}\}_{a,x}$ and ensemble $\{ {\sigma}_{AB}^{e e^\prime(0)} \}_{e e^\prime}$ by random. In order to get matrices $\left\{ M_{a|x}^{(0)} \geq 0\right\}_{a,x}$, we take $n_A m_A d_A^2$ real numbers between $-1$ and $1$ from a uniform distribution to write $n_A m_A$ tridiagonal matrices $\{T_{a|x}\}_{a,x}$. Then, we have $n_A m_A$ random hermitian and positive semi-definite matrices $\{M_{a|x}^{(0)} = T_{a|x}^{\dagger} T_{a|x} / {\rm Tr} [  T_{a|x}^{\dagger} T_{a|x}]  \}_{a,x}$. We note that the set of matrices may not be satisfied with $\sum_{a} M_{a|x}^{(0)} = \mathbb{I}_{d_A}$. But in the following steps, we set this condition as a constraint in SDP, then the feasible set is still constrained by conditions of POVM, i.e., the final optimal solution $\{M_{a|x}^{\rm opt}\}_{a,x}$ are POVMs. For the set of matrices $\{\sigma_{AB}^{e e^\prime(0)}\}_{e, e^\prime}$, we take $2 n_An_B d_A d_B$ real numbers between -1 and 1 from uniform distribution to generate the amplitude terms as well as phase terms of $n_An_B$ pure states $\{\rho_{AB}^{e e^\prime(0)}\}_{e,e^\prime}$. We also take $n_An_B$ positive numbers between 0 and 1 from uniform distribution to generate probability $p(e,e^\prime)$ with normalization $\sum_{e = 0}^{n_A - 1}\sum_{e^\prime=0}^{n_B-1} p(e,e^\prime)=1$. Hence, the random state $\sigma_{AB}^{e e^\prime(0)} = p(e,e^\prime)\rho_{AB}^{e e^\prime(0)}$ is obtained. After giving the initial POVMs $\{ M_{a|x}^{(0)} \}_{a,x}$ and ensemble $\{ \sigma_{AB}^{e e^\prime(0)} \}_{e, e^\prime}$, 
 we can solve the following SDP:
	\begin{eqnarray}
		\begin{aligned}
  \label{suppeq:4}
			\max \quad &\sum_{e,e^\prime}\operatorname{Tr}\left[  \left(  M_{a = e|x^*}^{(0)} \otimes M_{b = e^\prime|y^*} \right)  \sigma_{ AB}^{ee^\prime(0)} \right]  - \mu \lambda\\ 
   \text{w.r.t.} \quad & \{M_{b|y}\}_{b,y}, \lambda\\ 
   \operatorname{ s.t.} \quad& -\lambda \leq \sum_{a,b,x,y,e,e^\prime} I_{abxy}\operatorname{Tr}\left(M_{a|x}^{(0)} \otimes M_{b|y} \sigma_{AB}^{ee^\prime (0)}\right) - I^{\rm obs}  \leq \lambda,\\
			&\left\{M_{b|y}\right\}_b\in {\rm POVM}, \quad \forall y, \quad  \lambda\geq 0.
		\end{aligned}
	\end{eqnarray}
In this step, an optimal solution $\{M_{b|y}^{(1)}\}_{b,y}$ can be found easily.
Here we use a penalty to change the first equality constraint in Eq.~\eqref{suppeq:upperbound} to an inequality constraint, where $\mu$ is about 1e2$\sim$1e3. This is because the first constraint in Eq.~\eqref{suppeq:upperbound} can not be satisfied easily when the POVMs $\left\{ M_{a|x}^{(0)} \right\}_{a,x}$ and states $\left\{ \sigma_{AB}^{e e^\prime(0)} \right\}_{e, e^\prime}$ are set by random. Then Eq. \eqref{suppeq:upperbound} will probably be infeasible without penalty.

In the second step, we fix the optimal solution $\{M_{b|y}^{(1)}\}_{b,y}$ and also the initial  states $\{ \sigma_{AB}^{ee^\prime(0)} \}_{e, e^\prime}$. Then we can also find an optimal solution $\{M_{a|x}^{(1)}\}_{a,x}$ by changing the variables in Eq.~\eqref{suppeq:4}. Now the optimal solution $\{M_{a|x}^{(1)}\}_{a,x}$ is already a set of POVMs and can be easily found in this step. 

In the third step, we fix POVMs $\left\{M_{a|x}^{(1)}\right\}_{a,x}$ and $\left\{M_{b|y}^{(1)}\right\}_{b,y}$ to find an optimal set of states $\left\{ \sigma_{AB}^{ee^\prime(0)} \right\}_{e,e^\prime}$ that maximize Eq.~\eqref{suppeq:4}. Note that the third step will take much more time than the first two steps, so we iterate the first two steps until the guessing probability reaches its convergence. 

We iterate these steps until $\lambda$ decreases to about $10^{-9}$ and the optimal guessing probability reaches its convergency, hence an upper bound $H_{\min}^{\rm Dim}$ can be calculated. 

\subsection{Analytical upper bound of certifiable randomness in the maximal violation of the SATWAP inequality}
Firstly, the quantum bound of SATWAP inequality $I_{mn}^S$ can be achieved using the $d$-dimensional maximally entangled state $|\psi\rangle = \sum_{q=0}^{d-1} {1\over \sqrt{d}} |qq\rangle $ when Alice and Bob perform the measurements~\cite{prl_2017_SATWAP}:
\begin{equation}
    A_x = U_x^\dagger F\Omega F^\dagger U_x, \qquad B_y = V_y F^\dagger \Omega F V_y^\dagger,
\end{equation}
where $\Omega = \text{diag} \left[1,\omega,\omega^2,\cdots, \omega^{d-1}\right]$, with $\omega = \exp{\left( 2\pi i/d\right)}$, and $F$ is the $d\times d$ discrete Fourier transform matrix given by $F_d = {1\over \sqrt{d}} \sum_{i,j = 0}^{d-1} \omega^{ij} |i\rangle \langle j|$. Here $U_x$ and $V_y$ are 
\begin{equation}
    U_x = \sum_{j = 0}^{d-1} \omega^{j\theta_x} |j\rangle \langle j| , \qquad V_y = \sum_{j = 0}^{d - 1} \omega^{j \zeta_y} |j\rangle\langle j|
\end{equation}
with the phases $\theta_x = (x-1/2)/m$ and $\zeta_y = y/m$ for $x,y = 1,\cdots,m$. Thus, the quantum bound of SATWAP inequality can be achieved by the following probabilities
\begin{equation}
    \begin{aligned}
        P(A_x=a,B_y=b) = \left|{1\over d}\sum_{q=0}^{d-1} {1\over \sqrt{d}} \exp{\left({2\pi i\over d}q \left(a-b-\theta_x + \zeta_y\right)\right)}\right|^2.
    \end{aligned}
\end{equation}
 Since these probabilities only depend on the difference of $k = a-b$, we have 
\begin{equation}
    P(A_x=b+k,B_y=b) = P(A_x=k,B_y=0).
\end{equation}
We note that the arithmetic is taken modulo $d$ and $\exp \left(2\pi i q l d/d \right) = 1$ for $l\in \mathbb{Z}$.
Then, for the certain measurements $x^*,y^*$, the probability of most likely outcomes is
\begin{equation}
\label{eq:s16}
    \max_{a,b}P(A_{x^*} = a,B_{y^*} = b) = \max_k P(A_{x^*} = k, B_{y^*} = 0) = \max_k {1\over d^3}\left|\sum_{q=0}^{d-1} \exp{\left({2\pi i\over d}q \left(k+{y^*-x^*+{1\over 2}\over m}\right)\right)}\right|^2.
\end{equation}
Here we define $t = k+c$ and $c = {y^*-x^*+{1\over 2}\over m}\in[- 1 + 3/2m , 1-1/2m]$, $k\in\{0,1,\cdots,d-1\}$. We further simplify the right hand side of Eq.~\eqref{eq:s16} as
\begin{equation}
    \begin{aligned}
        {1\over d^3}\left|\sum_{q=0}^{d-1} \exp{\left({2\pi i\over d}q t\right)}\right|^2 = {1\over d^3}\left|\frac{\exp\left({2\pi i} t\right) - 1}{\exp\left({2\pi i} t/d\right) - 1}\right|^2 = {1\over d^3}\frac{1-\cos{(2\pi t)}}{1-\cos{(2\pi t/d)}}={1\over d^3}\left(\frac{\sin(\pi c)}{\sin(\pi (k+c)/d)}\right)^2.
    \end{aligned}
\end{equation}
Then, for any $c\in[- 1 + 3/2m , 1-1/2m]$, the right hand side of Eq.~\eqref{eq:s16} is
\begin{equation}
    \begin{aligned}
        \max_k {1\over d^3}\left(\frac{\sin(\pi c)}{\sin(\pi (k+c)/d)}\right)^2 = {1\over d^3} \max \left\{ \left(\frac{\sin(\pi c)}{\sin(\pi c/d)}\right)^2, \left(\frac{\sin(\pi c)}{\sin(\pi (c+1)/d)}\right)^2, \left(\frac{\sin(\pi c)}{\sin(\pi (c-1)/d)}\right)^2 \right\},
    \end{aligned}
\end{equation}
as $(k+c)/d \in (-1/d,1)$ and $(k+c)/d \in (0,1)$ when $k\geq 1$. 
Therefore, we define $f_1(c) = \left(\frac{\sin(\pi c)}{\sin(\pi c/d)}\right)^2$, $f_2(c) = \left(\frac{\sin(\pi c)}{\sin(\pi (c+1)/d)}\right)^2$ and $f_3(c) = \left(\frac{\sin(\pi c)}{\sin(\pi (c-1)/d)}\right)^2$, and we will compare these functions in the following.

Since $f_1(c)$ is an even function and $f_2(-c) = f_3(c)$, we only need to consider $c\in[0,1-1/2m]$. Moreover, we only have to consider $f_1(c)$ as well as $f_3(c)$ as $[\sin(\pi(c+1)/d)]^2 \geq [\sin(\pi(c-1)/d)]^2$ in this region. Then we show that $f_1(c)$ and $f_3(c)$ are monotone in the region of $c\in[0,1-1/2m]$:
\begin{equation}
    \begin{aligned}
        &f_1'(c) = \frac{2\pi \sin (\pi c)}{d \sin (\pi c/d)} \frac{d\cos(\pi c)\sin(\pi c/d) - \cos(\pi c/d)\sin(\pi c)}{\sin^2(\pi c /d)}\\ 
        &f_3'(c) = \frac{2\pi \sin (\pi c)}{d \sin (\pi (c-1)/d)} \frac{d\cos(\pi c)\sin(\pi (c-1)/d) - \cos(\pi (c-1)/d)\sin(\pi c)}{\sin^2(\pi (c-1)/d)}\\
        &[d\cos(\pi c)\sin(\pi c/d) - \cos(\pi c/d)\sin(\pi c)]' = (-d+1/d)\pi \sin(\pi c)\sin(\pi c/d)\leq 0\\
        &[d\cos(\pi c)\sin(\pi (c-1))/d) - \cos(\pi (c-1))/d)\sin(\pi c)]' = (-d+1/d)\pi \sin(\pi c)\sin(\pi (c-1)/d) \geq 0
    \end{aligned}
\end{equation}
We find $f_1'(c)\leq 0$ and $f_3'(c)\geq 0$ in the range of $c\in[0,1-1/2m]$. Since $f_1(1/2) = f_3(1/2)$, we have 
\begin{equation}
    \max_{a,b}P(A_{x^*} = a,B_{y^*} = b) = {1\over d^3}\max_{k} \left(\frac{\sin(\pi c)}{\sin(\pi (k+c)/d)}\right)^2 \geq  {1\over d^3} \frac{1}{\sin^2(\pi /2d)}
\end{equation}
The equality holds if and only if $c = 1/2$. Thus, we have 
\begin{equation}
    \min_{x^*,y^*}\max_{a,b}P(A_{x^*} = a,B_{y^*} = b) = {1\over d^3}\min_c\max_{k} \left(\frac{\sin(\pi c)}{\sin(\pi (k+c)/d)}\right)^2 \geq  {1\over d^3} \frac{1}{\sin^2(\pi /2d)}
\end{equation}
For odd $m$, $c = {y^*-x^*+{1\over 2}\over m}$ can achieve $1/2$ when $y^* - x^* = (m-1)/2$. For even $m$, we have
\begin{equation}
    \min_{x^*,y^*}\max_{a,b}P(A_{x^*} = a,B_{y^*} = b) = {1\over d^3} \min_c\max_{k} \left(\frac{\sin(\pi c)}{\sin(\pi (k+c)/d)}\right)^2 = {1\over d^3}\left(\frac{\sin[\pi (1/2-1/2m)]}{\sin[\pi (1/2-1/2m)/d]}\right)^2 > {1\over d^3} \frac{1}{\sin^2(\pi /2d)}.
\end{equation}
Here we note that $f_1(1/2 - x) = f_3(1/2 + x)$.

In summary, for odd $m$, when the SATWAP inequality is maximally violated, the upper bound of certifiable randomness is
\begin{equation}
     -\log_2\left( \min_{x^*,y^*}\max_{a,b}P(A_{x^*} = a,B_{y^*} = b) \right) = 3\log_2d + 2\log_2 \sin\left({\pi\over 2d}\right),
\end{equation}
which is independent with $m$ and the maximum randomness $2\log_2 d$ bits can be achieved only when $d=2$. 
For even $m$, the upper bound of certifiable randomness is
\begin{eqnarray}
    -\log_2\left( \min_{x^*,y^*}\max_{a,b}P(A_{x^*} = a,B_{y^*} = b) \right) = 3\log_2d + 2\log_2 \sin\left({\pi(1/2-1/2m)\over d}\right) - 2\log_2 \sin\left({\pi(1/2-1/2m)}\right).
\end{eqnarray}
For the inequalities with inputs $m\in\left\{ 2,3 \right\}$ and outputs $n=d\in\left\{ 2,4 \right\}$, the corresponding analytical upper bounds are $H_{\min} (m = 2, n = 2) \leq 1.228$ bits, $H_{\min} (m = 3, n=2) \leq 2.000$ bits, $H_{\min} (m = 2, n=4) \leq 2.284$ bits, $H_{\min} (m = 3, n=4) \leq 3.228$ bits. We find that the SATWAP inequalities with $m = 3$ inputs can potentially certify higher randomness compared with the case of $m = 2$ inputs. Additionally, these upper bounds are tight as the difference between these upper bounds and lower bounds (at the level $1+AB$ of the NPA hierarchy) are on the order of $10^{-3}$.

\subsection{Testing the structure of measurement incompatibility}
To check the structure of Alice's measurement incompatibility when a violation of SATWAP inequality is observed, we solve the following optimization problem~\cite{prl_2019_detectICstructure}:
\begin{equation}
    \begin{aligned}
        \label{eq:testICstructure}&\mathcal{B}_{mn}^A(x^*) = \text{max}_{\{p(ab|xy)\}} ~~ I_{mn}^{\text{S}}\left(p(ab|xy)\right)\\ 
        &\text{s.t.} ~~~~~p(ab|xy)\in L_{x^*\cap {\bar x^*}}^A\\
        & ~~~~~~~~p(ab|xy)\in Q_{1}
    \end{aligned}
\end{equation}
where $Q_{1}$ corresponds to the first level of NPA hierarchy, $L^A_{x^*\cap {\bar x^*}} \coloneqq \bigcap_{x\neq x^*} L_{x^*x}^A$ and $L_{x^*x}^A$ corresponds to the behaviors given when Alice's measurements $x^*$ and $x$ are compatible. A behavior belonging to $L_{x^*x}^A$ means it admits the LHV model when Alice's measurements are restricted to $\{x^*,x\}$. For example, the observation that $p(ab|xy)\notin L_{12}^A$ means the measurements $x=1$ and $x=2$ must be incompatible. Similarly, the maximally achievable violation within the same structure of Bob's measurements can also be obtained, denoted by $\mathcal{B}_{mn}^B$.

The numerical result is shown in Table~\ref{table:OnePartPairwise}. Here $\mathcal{C}_{mn}^{\text{S}}$ is the classical bound of SATWAP inequality, and we find the optimal value of this problem is consistent with the classical bound of SATWAP inequality (their difference coincide up to $10^{-9}$). It shows that violating these SATWAP inequalities means measurements $x^*$ and $y^*$ cannot be compatible with every other measurement respectively, i.e., $p(ab|xy)\notin L_{x^*\cap \bar{x}^*}$. Moreover, as the discussion shown above, the fact that the measurements $x^*$ and $y^*$ are incompatible with one of the rest measurements indicates nonzero randomness can be certified. Therefore, nonzero randomness can be certified after violating these SATWAP inequalities immediately.
\begin{table}[!ht]
    \centering
    \begin{tabular}{|c|c|c|c|}
    \hline
        $n$ & $m$ & $\max_{x^*}\mathcal{B}_{mn}^A (x^*) - \mathcal{C}_{mn}^{\text{S}}$ & $\max_{y^*} \mathcal{B}_{mn}^B (y^*) - \mathcal{C}_{mn}^{\text{S}}$ \\ \hline
        2 & 3 & $9.7920\times10^{-10}$ & $6.4615\times 10 ^{-12}$  \\ \hline
        4 & 3 & $4.2037\times 10^{-09}$ & $1.1922\times10^{-11}$ \\ \hline
    \end{tabular}
    \caption{\label{table:OnePartPairwise}Maximally achievable violation of SATWAP inequality within the constraint of $L_{x^*\cap {\bar x^*}}$.}
\end{table}

    Further, we note that the set $L_{x^*x}^A$ in Eq.~\eqref{eq:testICstructure} consisted of behaviors $\left\{ p(ab|xy)\right\}$ that is Bell local when Alice's measurements are restricted to $\{x^*,x\}$. Thus, verifying that a quantum behavior is outside of $L_{x^*x}^A$ is one way to certify that the measurements $x^*$ and $x$ are incompatible in a device-independent manner. 
    
    However, locality does not necessarily imply the measurement compatibility~\cite{prl_2019_detectICstructure}. To analyze the $I_{42}^{\rm F}$ inequality in the context of testing measurement incompatibility, we now consider the set $Q_{x^*x,y^*y}^{\rm JM}$. A behavior $p(ab|xy)$ belongs to $Q_{x^* x,y^* y}^{\rm JM}$ means the measurements $\{x^*,x\}$ ($\{y^*,y\}$) are substituted by a single parent measurement $\left\{M_{aa'}^{x^*x}\right\}_{aa'}$ ($\left\{M_{bb'}^{y^*y}\right\}_{bb'}$), i.e., $M_{a|x^*} = \sum_{a'}M_{aa'}^{x^*x}$ and $M_{a'|x} = \sum_{a}M_{aa'}^{x^*x}$, and likewise for Bob. In terms of the NPA hierarchy, this can be obtained by taking the moments involving $M_{aa'}^{x^*x}$ ($M_{bb'}^{y^*y}$) instead of $\left\{ M_{a|x^*}\right\}$ and $\left\{ M_{a|x}\right\}$ ($\left\{ M_{b|y^*}\right\}$ and $\left\{ M_{b|y}\right\}$)~\cite{prr_2021_quantifyIC,prl_2019_detectICstructure}. Then we solve the following optimization problem:
    \begin{equation}
        \begin{aligned}
            \mathcal{B}^{Q}(x^*,y^*) &= \max_{\{p(ab|xy)\}} I_{42}^F\left( p(ab|xy) \right) \\
            {\rm s.t. }~~ & p(ab|xy) \in \cap_{x\neq x^*, y\neq y^*} Q_{x^*x,y^*y}^{\rm JM},\\
            & p(ab|xy) \in Q_2,
        \end{aligned}
    \end{equation}
    We take the level $2$ of the NPA hierarchy and obtain $\min_{x^*,y^*} \mathcal{B}^{Q}(x^*,y^*) = 0.517$. When the observed Bell violation exceeds this bound, it means Alice’s measurements cannot be explained by the structure where $x^*$ is compatible with every rest measurement (same for Bob’s measurement $y^*$), which explains the nonzero randomness for $I_{42}^{\rm F} > 0.517$, although this bound is a little bit higher than the requirement for certifying randomness 0.499 as shown in Fig. 2 (a) in the main text. For Fig.~2 in the main text, we use the see-saw algorithm to determine the upper bound of guessing probability for $I_{42}^F \leq 0.499$, where we restrict the dimension of each subsystem to $6$ and identify examples giving the violation of $0.499$ and the guessing probability of $1$. In these examples, both the compatibility structures of Alice’s and Bob’s measurements conform to the specific compatibility structure, i.e., the measurements $x^*$ and $y^*$ are compatible with each of the rest measurements individually.

\subsection{Detection efficiency of certifiable randomness via Bell inequality}
To take into account inconclusive events we choose that Alice and Bob output ``$d-1$" in case of nondetection. When the detection efficiency of Alice and Bob is $\eta$, the probabilities are thus modified according to
\begin{equation}
    \begin{aligned}
        P^\eta(ab|xy) = 
        \begin{cases}
            \eta^2 P(ab|xy) & \forall x,y, a,b\in\{0,1,\cdots,d-2\},\\
            \eta^2 P(ab|xy) + \eta(1-\eta)P(a|x) & \forall x,y, a\in\{0,1,\cdots,d-2\},b=d-1,\\
            \eta^2 P(ab|xy) + \eta(1-\eta) P(b|y) & \forall x,y, a=d-1, b\in\{0,1,\cdots,d-2\},\\
            \eta^2 P(ab|xy) + (1-\eta)^2 + \eta (1-\eta)\left[P(b|y) + P(a|x)\right] & \forall x,y, a=d-1, b=d-1.
        \end{cases}
    \end{aligned}
\end{equation}
Therefore, for each $\eta$, the maximal value of a given inequality can be solved by the following problem:
\begin{equation}
    \begin{aligned}
       \mathcal{S} =  &\max_{\{P(ab|xy)\}} ~ \sum_{abxy}I_{abxy}P^\eta(ab|xy)\\
        &~{\rm s.t.} ~~~ P(ab|xy)\in Q.
    \end{aligned}
\end{equation}

\begin{figure*}[ht!]
\centering
\includegraphics[width=0.42\textwidth]{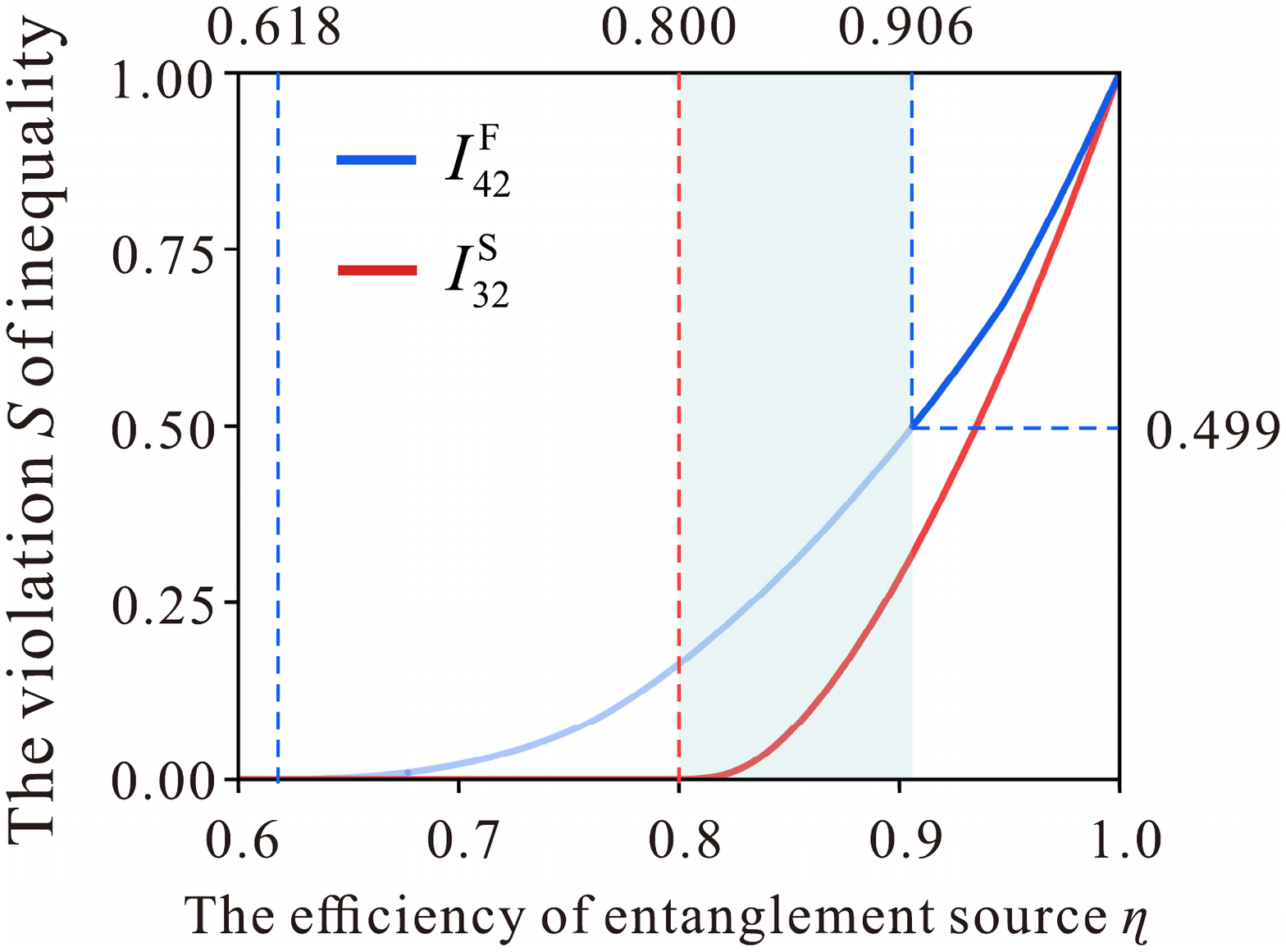}	\caption{\label{fig:detection efficiency} Maximal achievable value (normalized) $\mathcal{S}$ of Bell inequalities versus the detection efficiency $\eta$. The $I_{32}^\text{S}$ inequality (red curve) can tolerate the detection efficiency lower down to $80.0\%$. Due to its sensitivity in certifying randomness, the entire red curve corresponds to the case of nonzero randomness. Though the detection efficiency as low as $61.8\%$ is acceptable using the $I_{42}^\text{F}$ inequality (blue curve). However, to certify randomness, the $I_{42}^\text{F}$ inequality demands a violation of $0.499$, leading to a higher threshold of detection efficiency at $90.6\%$. Consequently, the scenario in the dark blue curve has Bell's nonlocal and nonzero randomness that can be certified, while the scenario in the light blue curve only has Bell's nonlocal.}
\end{figure*}

As shown in Fig.~\ref{fig:detection efficiency}, the maximal value $\mathcal{S}$ is close to the quantum bound of the given Bell expression when $\eta = 1$. For a detection efficiency lower than threshold $\eta_{\mathrm{th}}$, $\mathcal{S}$ is less than or equal to the classical bound of the expression. It means no violation is possible if $\eta\leq\eta_{\mathrm{th}}$. We find that $\eta_{\mathrm{th}}^\text{S} = 0.800$ for the SATWAP inequality with $3$ inputs and $2$ outputs in the level of $1+AB$ of NPA hierarchy, and $\eta_{\mathrm{th}}^{\text{F}} = 0.618$ for the facet inequality $I_{42}^{\text{F}}$ in the level of $1+AB+AAB$ of NPA hierarchy. To show the tightness of these bounds, we can find quantum behaviors to violate these inequalities by using the see-saw algorithm when the detection efficiency is $\eta_{\mathrm{th}} + 0.001$. Due to the nonzero randomness requires the violation of the facet inequality $I_{42}^{\text{F}}$ must exceed 0.499, it can only tolerate the detection efficiency with $\eta^\text{F}>90.6\%$ to generate randomness. Therefore, one can certify randomness based on the SATWAP inequality within a broader region of detection efficiency (light blue area).

\section{Quantum states preparation and corresponding projection measurements under two types of inequalities}
\begin{figure*}[ht!]
\centering
\includegraphics [width=1\textwidth]{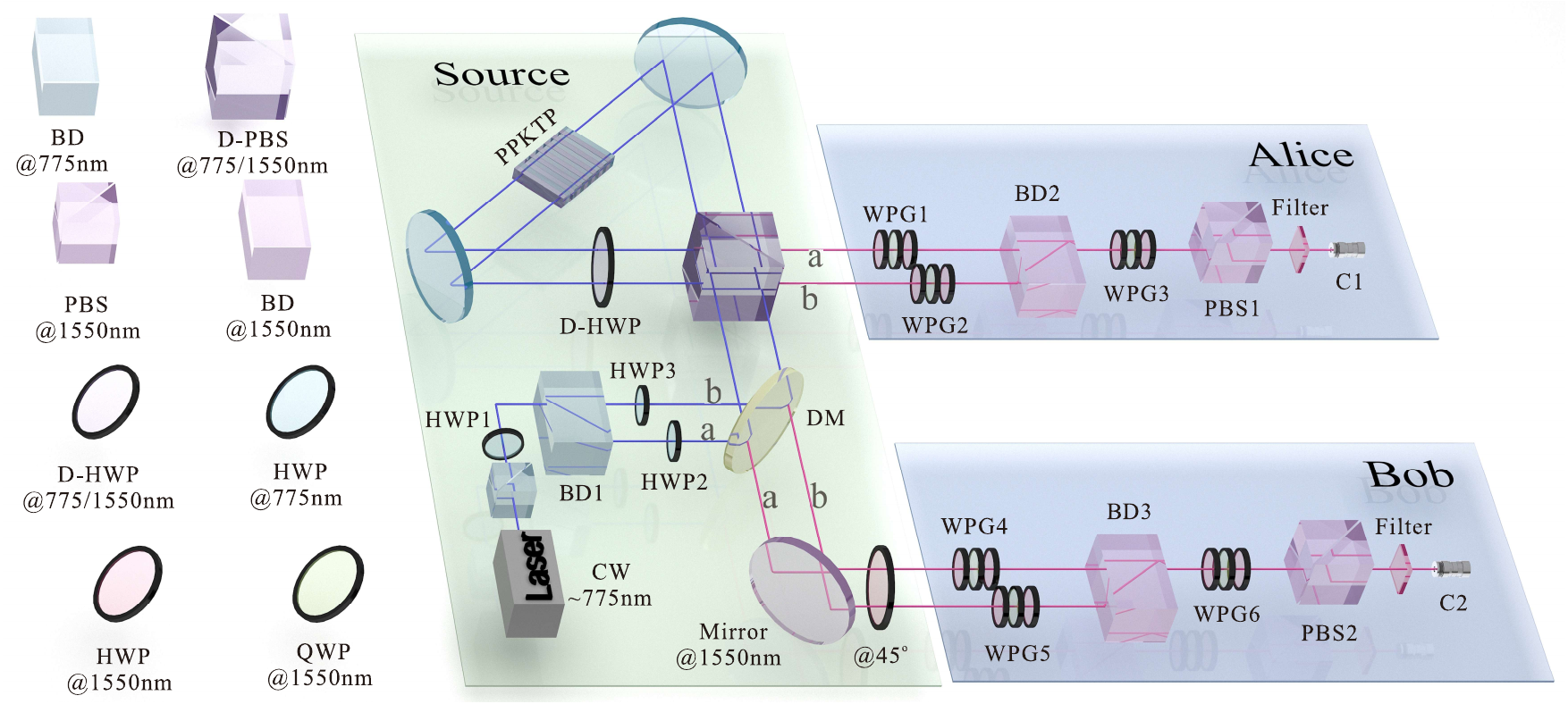}
\vspace{-0.3cm}%
\caption{Experiment setup. Prepare corresponding quantum states, then distribute entangled photon pairs to Alice and Bob for local projection measurements, respectively.}
\label{fig:setup}
\vspace{-0.2cm}%
\end{figure*}

In Figure \ref{fig:setup}, the continuous wave (CW) with a wavelength of 775 nm is horizontally polarized by PBS (@775 nm), and after passing through HWP1 with angle $\frac{\theta_1}{2}$ and BD1, which separates parallel paths (a, b) of horizontally and vertically polarized photons with a spacing of 4mm. Its state is $|\Psi\rangle=\cos{\theta_1}|a_H\rangle+\sin{\theta_1}|b_V\rangle$, where $|a(b)_{H(V)}\rangle$ represents photons in path a(b) with horizontal(vertical) polarization. After passing through HWP2 and HWP3 with the angles $\frac{\theta_2}{2}$ and $\frac{\pi+2\theta_3}{4}$ respectively, and the status is $|\Psi\rangle=\cos (\theta_1)\cos (\theta_2)|a_H\rangle+\cos(\theta_1)\sin (\theta_2) |a_V\rangle+\sin (\theta_1)\cos (\theta_3)|b_H\rangle+\sin (\theta_1)\sin (\theta_3)|b_V\rangle$. Choose polarization and path degrees of freedom (DOF) for encoding, where $|a_H\rangle$, $|a_V\rangle$, $|b_H\rangle$, and $|b_V\rangle$ are encoded as $|0\rangle$, $|1\rangle$, $|2\rangle$, and $|3\rangle$, respectively. The spontaneous parametric down-conversion (SPDC) process occurs on PPKTP in a Sagnac structure, generating the quantum entangled state 
\begin{equation}
    |\Psi\rangle=\cos (\theta_1)\cos (\theta_2)|00\rangle+\cos(\theta_1)\sin (\theta_2) |11\rangle+\sin (\theta_1)\cos (\theta_3)|22\rangle+\sin (\theta_1)\sin (\theta_3)|33\rangle
\end{equation}
Tilt BD1, HWP2, and HWP3 can adjust the relative phase between each subspace, so we can prepare any entangled state up to four dimensions.  

Distribute entangled photons to Alice and Bob for local projection measurements, respectively. The measurement of high-dimensional projection can be decomposed into projections on a two-dimensional subspace. The wave plate group (WPG) consists of a set of wave plates (QWP, HWP, QWP) that can perform any operation in two-dimensional space. WPG1 and WPG2, in conjunction with BD2, can achieve projection measurement $cos{\alpha_1}|0\rangle+e^{i\phi_1}\sin{\alpha_1}|1\rangle$ of subspaces $\{|0\rangle, |1\rangle\}$, and $\cos{\alpha_2}|2\rangle+e^{i\phi_2}\sin{\alpha_2}|3\rangle$ of subspaces $\{|2\rangle, |3\rangle\}$. Subsequently, WPG3, in conjunction with PBS1, can complete any projection measurement in four-dimensional space $\cos{\alpha_3}(\cos{\alpha_1}|0\rangle+e^{i\phi_1}\sin{\alpha_1}|1\rangle)+e^{i\phi_3}\sin{\alpha_3}(\cos{\alpha_2}|2\rangle+e^{i\phi_2}\sin{\alpha_2}|3\rangle)$.

Importantly, when measuring two-dimensional subspaces using the computational basis $\{|0\rangle, |1\rangle\}$ and Fourier basis $\{(|0\rangle+|1\rangle)/\sqrt{2}, (|0\rangle-|1\rangle)/\sqrt{2}\}$, the corresponding visibility are $0.999\pm0.001$ and $0.990\pm0.001$, respectively. Therefore, we can accurately prepare the target entangled state.

We consider the following scenario. Alice and Bob share a pair of entangled states and can choose from the $m$ measurement settings, each with $n$ output. The measurement setting selected by Alice is denoted as $A_x$ with $x\in\{1,2,...,m\}$ and her output is denoted as $a\in\{1,2,...,n\}$. Similarly, Bob has $B_y$ with $y\in\{1,2,...,m\}$ and output $b\in\{1,2,...,n\}$. In our experiment, we characterized a set of joint probabilities. The $P(A_x^a,B_y^b)$ represents the probability of obtaining the output $a$ and $b$, when Alice and Bob perform the measurement setting $A_x$ and $B_y$, respectively. Additionally, $P(A_x^a)$ (or $P(B_y^b))$ represents the probability of obtaining the output $a$ (or $b$) when Alice (or Bob) performs the measurement setting $A_x$ (or $B_y$) and Bob (or Alice) do noting. Next, utilizing the facet Bell inequalities $I_{mn}^\text{F}$ and the SATWAP inequalities $I_{mn}^{\text{S}}$, we prepare entangled states and construct projection measurements tailored to these specific inequalities to test the Bell nonlocality.

When testing the Bell nonlocality of entangled states using inequalities $I_{22}^\text{F}$ and $I_{42}^\text{F}$, and considering measurements $A_x$ and $B_y$ with outputs $a$ and $b$, we conducted experiments by recording events for a duration of 30 seconds, denoted as $C(A_x^a,B_y^b)$. In cases where one party, either Alice or Bob, performed a measurement with a specific outcome (e.g., $A_x$ with outcome $a$ or $B_y$ with outcome $b$) while the other did nothing, the nonactive party's measurement corresponds to the identity matrix. This identity matrix can be expressed as a sum over a set of complete measurement bases ${E_i}$, such as computational bases ${|1\rangle, ..., |d\rangle}$, where $d$ represents the dimension of the entangled state. As a result, the probability $P(A_x^a)$ (or $P(B_y^b)$) is given by the sum of the probabilities over all the measurement bases: $P(A_x^a) = \sum_{i=1}^{d} P(A_x^a, E_i)$ (or $P(B_y^b) = \sum_{i=1}^{d} P(E_i, B_y^b)$). Subsequently, we determined the corresponding event counts, denoted as $C(A_x^a)$ (or $C(B_y^b)$). The total number of events $N$ was estimated based on the statistical time, with the source generating 5000 pairs of entangled photons per second. In our specific experimental setup, the total number of events was approximately $N = 150,000$. Therefore, the normalized probabilities were calculated as follows: $P(A_x^a, B_y^b) = C(A_x^a, B_y^b)/N$, $P(A_x^a) = C(A_x^a)/N$, and $P(B_y^b) = C(B_y^b)/N$. When employing other inequalities such as $I_{22}^\text{F}$ and $I_{mn}^{\text{S}}$ to test the Bell nonlocality of entangled states, the event count $C(A_x^a, B_y^b)$ was determined over a 30-second interval, and the probabilities were normalized as $P(A_x^a, B_y^b) = C(A_x^a, B_y^b)/\sum\sum_{a,b=1}^{n}C(A_x^a, B_y^b)$ for measurements $A_x$ and $B_y$ with outputs $a$ and $b$. In summary, by employing specific Bell inequalities, entangled states, and projection measurements, we experimentally tested the Bell nonlocality of entangled states.

\subsection{The facet Bell inequalities $I_{mn}^{\text{F}}$.}

For the facet Bell inequalities $I_{mn}^{\text{F}}$, we choose $(m, n)=(2,2), (2,4)$, and $(4,2)$, which are inequalities $I_{22}^\text{F}$, $I_{24}^\text{F}$, and $I_{42}^\text{F}$. For each inequality, we select several entangled states $|\Psi_k^{mn}\rangle$ and construct corresponding projection measurements for Bell nonlocality testing, where $|\Psi_k^{mn}\rangle$ represents the $k$th entangled state selected for inequality $I_{mn}^{\text{F}}$. Below is a detailed introduction to each inequality of $I_{mn}^{\text{F}}$.

\subsubsection{$I_{22}^\text{F}$ inequality with $(m,n)=(2,2)$}
When $(m,n)=(2,2)$, the Bell inequality $I_{22}^\text{F}\leq0$ in the local hidden variable (LHV) model, where
\begin{equation}
I_{22}^\text{F}=P\left(A_1^1, B_1^1\right)+P\left(A_2^1, B_1^1\right)+P\left(A_1^1, B_2^1\right)-P\left(A_2^1, B_2^1\right)-P\left(A_1^1\right)-P\left(B_1^1\right).
\end{equation}
Use inequality $I_{22}^\text{F}$ to test the Bell nonlocality of the five entangled states. The corresponding entangled states and measurements are as shown in TABLE ~\ref{I2222}.

\subsubsection{$I_{24}^\text{F}$ inequality with $(m,n)=(2,4)$}
When $(m,n)=(2,4)$, the Bell inequality $I_{24}^\text{F}\leq0$ in the local latent variable (LHV) model, where
\begin{equation}
\begin{aligned}
I_{24}^\text{F} \equiv \sum_{k=0}^{[d / 2]-1}\left(1-\frac{2 k}{d-1}\right) & \left\{+\left[P\left(A_1=B_1+k\right)+P\left(B_2=A_1+k+1\right)+P\left(A_2=B_2+k\right)+P\left(B_1=A_2+k\right)\right]\right. \\
& -\left[P\left(A_1=B_1-k-1\right)+P\left(B_2=A_1-k\right)+P\left(A_2=B_2-k-1\right)\right. \left.\left.+P\left(B_1=A_2-k-1\right)\right]\right\} .
\end{aligned}
\end{equation}

Here, $P(A_x=B_y+s)=\sum_{r=1}^{n}P(A_x^{mod\{r-1+s,n\}},B_y^r)$ and $d$ is the dimension of entangled states. Use inequality $I_{24}^\text{F}$ to test the Bell nonlocality of the five entangled states. The corresponding entangled states and measurements are as shown in TABLE ~\ref{I2244}.

\subsubsection{$I_{42}^\text{F}$ inequality with $(m,n)=(4,2)$}
When $(m,n)=(4,2)$, the Bell inequality $I_{42}^\text{F}\leq0$ in the local latent variable (LHV) model, where
\begin{equation}
\begin{aligned}
I_{42}^\text{F}= & I_{\mathrm{CH}}^{(1,2 ; 1,2)}+I_{\mathrm{CH}}^{(3,4 ; 3,4)}-I_{\mathrm{CH}}^{(2,1 ; 4,3)}-I_{\mathrm{CH}}^{(4,3 ; 2,1)}-P\left(A_2^1\right)-P\left(A_4^1\right)-P\left(B_2^1\right)-P\left(B_4^1\right).
\end{aligned}
\end{equation}
Here, $I_{\mathrm{CH}}^{(i, j ; p, q)} \equiv P\left(A_i^1, B_p^1\right)+P\left(A_j^1, B_p^1\right)+P\left(A_i^1, B_q^1\right)-$ $P\left(A_j^1, B_q^1\right)-P\left(A_i^1\right)-P\left(B_p^1\right)$. Use inequality $I_{42}^\text{F}$ to test the Bell nonlocality of the seven entangled states. The corresponding entangled states and measurements are as shown in TABLE ~\ref{I4422}.

So far, we have demonstrated the entangled states and projection measurements for inequalities $I_{mn}^{\text{F}}$ used in the experiment. Prepare the entangled states mentioned above in our experimental system and perform corresponding projection measurements. The values obtained experimentally and the corresponding ideal values of the facet Bell inequalities $I_{mn}^{\text{F}}$ along with certifiable randomness are presented in Table \ref{Immnn_theory_and_exp} and Figure \ref{theory and exp tu}.(a). 
\begin{table}
    \centering
    \renewcommand\arraystretch{1.22}
    \begin{tabular}{|c|c|c|c|c|c|c|c|c|c|c|}
\hline \;$m$\; & \;$n$\; & $I_{mn}^{\text{F}}$ & state & violation(ideal) & violation(exp) &error & $H_{\min}(\overline{I})$(ideal)  & $H_{\min}(\overline{I}-\sigma)$ & $H_{\min}(\overline{I})$(exp) & $H_{\min}(\overline{I}+\sigma)$ \\
\hline 
& & & $\Psi_1^{22}$ &  0.2415 &  0.1989 &  0.0121 & 0.0959 &  0.0705 &  0.0759  &  0.0815 \\
& & & $\Psi_2^{22}$ &  0.4831 &  0.4396 &  0.0133 &  0.2355 &  0.1984 &  0.2068  &  0.2154\\
2 & 2 & {$I_{22}^\text{F}$}& $\Psi_3^{22}$ &  0.7246 &  0.6791 & 0.0181 &  0.4456 &  0.3799 &  0.3977 &  0.4163 \\
& & & $\Psi_4^{22}$ & 0.9662 &  0.9167 &  0.0186 &  0.9105 &   0.7110 &  0.7545 &  0.8046 \\
& & & $\Psi_5^{22}$ &  1.0000 &  0.9466 &  0.0196 &  1.1865 &   0.7815 &  0.8397 &  0.9126 \\
\hline 
& & & $\Psi_1^{24}$ &  0.3019 &  0.2665 &  0.0041 &  0.2152 &  0.1808 &  0.1843 &  0.1878 \\
& & & $\Psi_2^{24}$ &  0.6159 &  0.5995 &  0.0036 & 0.5805 &  0.5509 &  0.5563 &  0.5616 \\
2 & 4 & ${I^\text{F}_{24}}$ & $\Psi_3^{24}$ &  0.8454 &  0.8242 &  0.0036 & 1.0667 &   0.9951 & 1.0053 & 1.0156\\
& & & $\Psi_4^{24}$ &  0.9662 &  0.9184 &  0.0031 & 1.6511 &  1.3339 & 1.3491 & 1.3647 \\
& & & $\Psi_5^{24}$ &  1.0000 &  0.9623 &  0.0029 & 2.2532 & 1.5993 & 1.6226 & 1.6470\\
\hline 
& & & $\Psi_1^{42}$ &  0.2415 &  0.2064 &  0.0068 & 0.0000 &  0.0000 &  0.0000 &  0.0000 \\
& & & $\Psi_2^{42}$ &  0.3623 &  0.3119 &  0.0084 & 0.0000 &  0.0000 &  0.0000 &  0.0000 \\
& & & $\Psi_3^{42}$ &  0.4831 &  0.4447 &  0.0123 & 0.0000 &  0.0000 &  0.0000 &  0.0000 \\
4 & 2 & ${I_{42}^\text{F}}$ & $\Psi_4^{42}$ &  0.5435 &  0.5104 &  0.0099 &  0.0534 &  0.0006 &  0.0126 &  0.0248 \\
& & & $\Psi_5^{42}$ &  0.6039 &  0.5794 &  0.0109 &  0.1314 &  0.0856 &  0.0997 &  0.1139 \\
& & & $\Psi_6^{42}$ &  0.8454 &  0.8022 &  0.0118 &  0.4958 &  0.4038 &  0.4229 &  0.4423 \\
& & & $\Psi_7^{42}$ &  1.0000 &  0.9604 &  0.0130 &  1.1842 &  0.8194 &  0.8743 &  0.9336 \\
\hline
\end{tabular}
    \caption{Select inequalities $I_{22}^\text{F}$, $I_{24}^\text{F}$, and $I_{42}^\text{F}$ to test the nonlocality of several quantum states, where $\Psi_k^{mn}$ represents the $k$th entangled state for the inequality $I_{mn}^{\text{F}}$ used in the experiment. Violations observed in the experiments and the certified randomness are presented in columns three and five, respectively, followed by the corresponding errors.}
    \label{Immnn_theory_and_exp}
\end{table}

\subsection{SATWAP inequalities $I_{mn}^{\text{S}}$}
For $m$ measurement settings with $n$ outputs, the corresponding SATWAP inequality $I^S_{mn}$ takes the following form:

\begin{equation}
    I_{mn}^{\text{S}}:=\sum_{k=0}^{\lfloor n / 2\rfloor-1}\left(\alpha_k \mathbb{P}_k-\beta_k \mathbb{Q}_k\right),
\end{equation}
where $\mathbb{P}_k:=\sum_{i=1}^m\left[P\left(A_i=B_i+k\right)+P\left(B_i=A_{i+1}+k\right)\right]$, $\mathbb{Q}_k:=\sum_{i=1}^m\left[P\left(A_i=B_i-k-1\right)+P\left(B_i=A_{i+1}-k-1\right)\right]$ with $A_{m+1}:=A_1+1$. Here the probability $P(A_i = B_i + k)$ means that the outcomes of measurement $A_i$ and $B_i$ differ modulo $d$ by $k$. The parameters $\alpha_k$ and $\beta_k$ are
$$
\begin{aligned}
& \alpha_k=\frac{1}{2 n} \tan \left(\frac{\pi}{2 m}\right)\left[g(k)-g\left(\left\lfloor\frac{n}{2}\right\rfloor\right)\right], \\
& \beta_k=\frac{1}{2 n} \tan \left(\frac{\pi}{2 m}\right)\left[g\left(k+1-\frac{1}{m}\right)+g\left(\left\lfloor\frac{n}{2}\right\rfloor\right)\right]
\end{aligned}
$$
with $g(x):=\cot (\pi(x+1 / 2 m) / n)$. For SATWAP inequalities $I_{mn}^{\text{S}}$, we choose $(m, n)=(2,2), (3,2), (2,4)$, and $(3,4)$, which are inequalities $I_{22}^\text{S}$, $I_{32}^\text{S}, I_{24}^\text{S}$, and $I_{34}^\text{S}$. For each inequality, we select several entangled states $|\Phi_k^{mn}\rangle$ and construct corresponding projection measurements for Bell nonlocality testing, where $|\Phi_k^{mn}\rangle$ represents the $k$th entangled state selected for inequality $I_{mn}^{\text{S}}$. Below is a detailed introduction to each inequality of $I_{mn}^{\text{S}}$.

For inequality $I_{22}^\text{S}$, i.e., $(m,n)=(2,2)$, we choose three entangled states and the same projection measurement, as shown in TABLE ~\ref{I22}.

For inequality $I_{32}^\text{S}$, i.e., $(m,n)=(3,2)$, we choose three entangled states and the same projection measurement, as shown in TABLE ~\ref{I32}.

For inequality $I_{24}^\text{S}$, i.e., $(m,n)=(2,4)$, we choose three entangled states and the same projection measurement, as shown in TABLE ~\ref{I24}.

For inequality $I_{34}^\text{S}$, i.e., $(m,n)=(2,4)$, we choose three entangled states and the same projection measurement, as shown in TABLE ~\ref{I34}.

\begin{figure}[h]
\centering
\vspace{-1mm}
\includegraphics[width=1\textwidth]{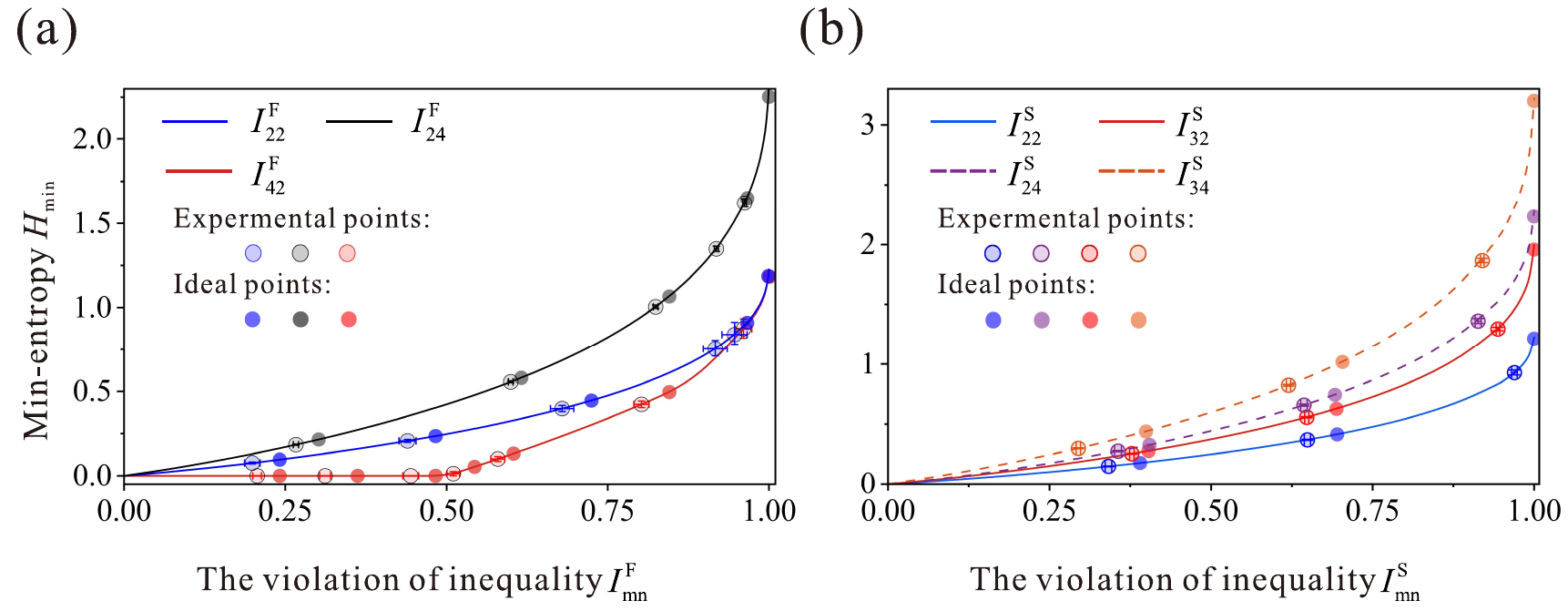}
    \caption{The horizontal axis represents the violation of inequality $I_{mn}^{\text{F}}$ (or $I_{mn}^{\text{S}}$) (normalized), and the vertical axis represents the lower bounds of certifiable randomness that the system can certify under the premise of testing Bell nonlocality. Experimental points in Figures (a) and (b) respectively show the violations of $I_{22}^{\text{F}}$, $I_{24}^{\text{F}}$, and $I_{42}^{\text{F}}$ of the facet inequality $I_{mn}^{\text{F}}$, and $I_{22}^{\text{S}}$, $I_{32}^{\text{S}}$, $I_{24}^{\text{S}}$ and $I_{34}^{\text{S}}$ of the SATWAP inequality obtained from experiments, and corresponding certifiable randomness. Ideal points in Figures (a) and (b) show the theoretical violations of $I_{22}^{\text{F}}$, $I_{24}^{\text{F}}$, and $I_{42}^{\text{F}}$ of the facet inequality $I_{mn}^{\text{F}}$, and $I_{22}^{\text{S}}$, $I_{32}^{\text{S}}$, $I_{24}^{\text{S}}$ and $I_{34}^{\text{S}}$ of the SATWAP inequality for given quantum states and measurements, and corresponding certifiable randomness.  For all the inequalities except the $I_{42}^{\text{F}}$ inequality, the lower bounds of the certifiable randomness were numerically determined at the level $1+AB$ of the NPA hierarchy. For the $I_{42}^{\text{F}}$ inequality, the certifiable randomness is bounded at the level of $1+AB+AAB$. We note that its upper bound is zero (up to numerical precision of $10^{-9}$) when $I_{42}^{\text{F}}\in [0,0.499]$. The error of Bell value is calculated by simulating photons with 1000 Poisson distribution. The lower (higer) certifiable randomness is calculated from the lower (higer) of Bell values, corresponding to its $-(+)1\sigma$ confidence interval.}
\vspace{-5mm} 
\label{theory and exp tu}
\end{figure}

So far, we have demonstrated the entangled states and projection measurements for SATWAP inequalities used in the experiment. Prepare the entangled states mentioned above in our experimental system and perform corresponding projection measurements. The values obtained experimentally and the corresponding ideal values of the SATWAP inequalities $I_{mn}^{\text{S}}$ along with certifiable randomness are presented in Table \ref{SATWAP_theort_and_exp} and Figure \ref{theory and exp tu}.(b).

\begin{table}[h]
    \centering
    \renewcommand\arraystretch{1.22}
    \begin{tabular}{|c|c|c|c|c|c|c|c|c|c|c|}
\hline \;$\mathrm{n}$\; & \;$\mathrm{m}$\; & $I_{mn}^{\text{S}}$ & state & violation(ideal) & violation(exp) & error &$H_{\min} (\overline{I} )$(ideal) &  $H_{\min} (\overline{I} - \sigma)$ &$H_{\min} (\overline{I} )$(exp)& $H_{\min} (\overline{I} + \sigma)$\\
\hline  &  &  & $\Phi_1^{22}$ & 1.0000 & 0.9696 &    0.0044 & 1.2158  &  0.9082 &  0.9276 &  0.9488 \\
 & 2 & $I_{22}^\text{S}$ & $\Phi_2^{22}$ & 0.6951 & 0.6490 &  0.0048 &  0.4140 &  0.3640 &  0.3685 &  0.3730 \\
2 & & & $\Phi_3^{22}$ & 0.3902 & 0.3414 &  0.0051 &  0.1762 &  0.1452 &  0.1480 &  0.1509 \\
\cline{2-11} &  &  & $\Phi_1^{32}$ & 1.0000 & 0.9440 &  0.0027 & 1.9576 & 1.2819 & 1.2960 & 1.3105 \\
 & 3 & $I_{32}^\text{S}$ & $\Phi_2^{32}$ & 0.6940 & 0.6480 &  0.0032 & 0.6263 &  0.5516 &  0.5562 &  0.5608 \\
 & & & $\Phi_3^{32}$ & 0.4033 & 0.3775 &   0.0035 & 0.2762 &  0.2499 &  0.2530 &  0.2561 \\
\hline  &  & &$\Phi_1^{24}$ & 1.0000 & 0.9133 &  0.0046 & 2.2339 & 1.3418 & 1.3624 & 1.3838 \\
 & 2 & $I_{24}^\text{S}$ & $\Phi_2^{24}$ & 0.6917 & 0.6436 & 0.0050 & 0.7433 &  0.6481 &  0.6565 &  0.6651 \\
4 & & & $\Phi_3^{24}$ & 0.4048 & 0.3561 &  0.0053 & 0.3267 &  0.2687 &  0.2742 &  0.2797\\
\cline{2-11} &  & & $\Phi_1^{34}$ & 1.0000 & 0.9198 &  0.0027 & 3.2000 & 1.8486 & 1.8665    & 1.8847\\
 & 3 & $I_{34}^\text{S}$ & $\Phi_2^{34}$ & 0.7032 & 0.6197 &   0.0034 &  1.0174 &  0.8152 &  0.8224 &  0.8296 \\
 & & & $\Phi_3^{34}$ & 0.3994 & 0.2947 &  0.0034 & 0.4386 & 0.2939 &  0.2981 &  0.3023 \\
\hline
\end{tabular}
    \caption{Select $m=2,3$ and $n=2,4$ for the SATWAP inequality $I_{mn}^{\text{S}}$ to test the nonlocality of several quantum states, where $\Phi_k^{mn}$ represents the $k$th entangled state for the inequality $I_{mn}^{\text{S}}$ used in the experiment. Violations observed in the experiments and the certified randomness are presented in columns four and six, respectively, followed by the corresponding errors.}
    \label{SATWAP_theort_and_exp}
\end{table}

\section{SCHEME FOR LOOPHOLE-FREE HIGH-DIMENSIONAL BELL TEST}

Here, we show how to implement a loophole-free Bell test using high-dimensional entanglement encoded with path-polarization degrees of freedom. In Bell testing, imperfections in experiments often create various loopholes that a local hidden variable (LHV) model can exploit to predict results. Besides the detection loophole discussed in the main text, there are two main loopholes in Bell tests: the locality (or communication) loophole and the freedom-of-choice loophole. To close the locality loophole, it is necessary to ensure that Alice and Bob cannot communicate with each other when they choose the measurement basis and conduct a measurement. To close the freedom-of-choice loophole, it is essential to ensure that the generation of entangled photon pairs does not affect Alice's or Bob's choice of measurement basis. Additionally, Alice and Bob's choices must be random. Therefore, it is crucial to ensure the space-like separation of Alice and Bob when selecting the basis and conducting measurements. Simultaneously, we need to ensure the space-like separation between Alice (or Bob) and the source of entanglement generation.

\begin{figure}[h]
\centering
\vspace{-1mm}
\includegraphics[width=1\textwidth]{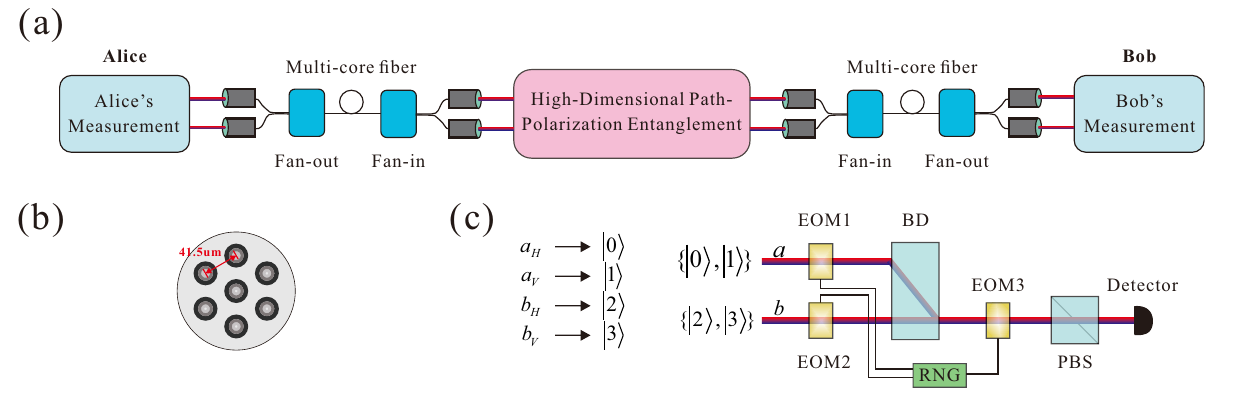}
    \caption{A scheme for a high dimensional loophole-free Bell test. The different colors on the paths represent different photon polarizations, indicating the encoding of two subspaces. (a) Distribution of high-dimensional entanglement. For high-dimensional entangled photon pairs with telecom wavelengths, multi-core fibers can be used for high-quality, long-distance distribution. (b) Schematic diagram of a multi-core optical fiber. (c) High dimensional projection measurement with high-speed switching. The different polarizations of paths $a$ and $b$ are encoded as ${|0\rangle, |1\rangle}$, and ${|2\rangle, |3\rangle}$, respectively. By utilizing a random number generator (RNG) to rapidly modulate the voltage of EOM1-3, it is possible to randomly switch between different projection measurements.}
\vspace{-2mm} 
\label{loophole-free}
\end{figure}


As shown in Fig.\ref{loophole-free}(a), multi-core optical fibers~\cite{Hu:20} can be used to distribute high-dimensional entangled states to locations Alice and Bob with high quality. A multi-core fiber consists of multiple single-mode fibers positioned very close to each other, providing a very stable structure. Additionally, efficient entangled states at telecom wavelength have been reported~\cite{prl_2022_belltest}, which can be distributed through optical fibers with low loss. High-speed measurement base switching is also crucial for achieving space-like separation. High-speed polarization rotation devices, such as electro-optic modulators (EOMs) as shown in Fig.\ref{loophole-free}(c), can replace half-wave plates to construct fast high-dimensional measurement devices. Finally, a high-speed random number generator (RNG) can drive the high-speed measurement devices to measure the distributed entangled states, enabling a loophole-free Bell test. The deployment of multi-core optical fibers and the implementation of rapid operations can introduce noise and loss, which diminish the fidelity and efficiency of entanglement sources. Addressing these challenges necessitates ongoing technological advancements.

\clearpage

\begin{table}[h]
    \centering
    \renewcommand\arraystretch{1.22}
    \begin{tabular}{|c|c|}
\hline 
Order & Quantum state and its projection measurements for inequality $I_{22}^\text{F}$ \\
\hline        
 & $|\Psi_1^{22}\rangle=-0.6965|00\rangle+0.3692|11\rangle+0.3855|22\rangle+0.4796|33\rangle$ \\
\cline{2-2}
1 & 
        $A_{1}^1=\left(\begin{array}{l}
           \,\:\; 0.2325 \\
           \,\:\; 0.6579 \\
           \,\:\; 0.5492 \\
           \,\:\; 0.4600
        \end{array}\right)$, $A_{2}^1=\left(\begin{array}{l}
           \,\:\; 0.5734 \\
           \,\:\; 0.4650 \\
           \,\:\; 0.4751 \\
           \,\:\; 0.4788
        \end{array}\right)$, $B_{1}^1=\left(\begin{array}{l}
            -0.4770 \\
            \,\:\;0.5851 \\
            \,\:\;0.5252 \\
            \,\:\;0.3927 
        \end{array}\right)$, $B_{2}^1=\left(\begin{array}{l}
           \,\:\; 0.4622 \\
           \,\:\; 0.5174 \\
           \,\:\; 0.5097 \\
           \,\:\; 0.5088
        \end{array}\right)$	 \\
\hline 
 & $|\Psi_2^{22}\rangle=-0.5711|00\rangle+0.4740|11\rangle+0.4739|22\rangle+0.4739|33\rangle $\\
\cline{2-2}
2 &  
        $A_{1}^1=\left(\begin{array}{l}
           \,\:\; 0.1823 \\
           \,\:\; 0.5677 \\
           \,\:\; 0.5676 \\
           \,\:\; 0.5676
        \end{array}\right)$, $A_{2}^1=\left(\begin{array}{l}
           \,\:\; 0.6722 \\
           \,\:\; 0.4275 \\
           \,\:\; 0.4274 \\
           \,\:\; 0.4274 \\
        \end{array}\right)$, $B_{1}^1=\left(\begin{array}{l}
            -0.4224 \\
            \,\:\;0.5233 \\
            \,\:\;0.5233 \\
            \,\:\;0.5233
        \end{array}\right)$,  $B_{2}^1=\left(\begin{array}{l}
           \,\:\; 0.3323 \\
           \,\:\; 0.5446 \\
           \,\:\; 0.5445 \\
           \,\:\; 0.5445
        \end{array}\right)$ \\
\hline
 & 
$|\Psi_3^{22}\rangle=-0.6761|00\rangle-0.7053|11\rangle-0.1526|22\rangle+0.1488|33\rangle $\\
\cline{2-2}
3 & 
        $A_{1}^1=\left(\begin{array}{l}
            \,\:\;0.2357 \\
            -0.9329 \\
            -0.1011 \\
            \,\:\;0.2529
        \end{array}\right)$, $A_{2}^1=\left(\begin{array}{l}
            \,\:\;0.8944 \\
            -0.4228 \\
            -0.0612 \\
            \,\:\;0.1324
        \end{array}\right)$, $B_{1}^1=\left(\begin{array}{l}
            -0.6711 \\
            \,\:\;0.7245 \\
            -0.0667 \\
            \,\:\;0.1428
        \end{array}\right)$,  $B_{2}^1=\left(\begin{array}{l}
            \,\:\;0.2027 \\
            \,\:\;0.9724 \\
            -0.0916 \\
            \,\:\;0.0702
        \end{array}\right)$ \\
\hline
 & $
|\Psi_4^{22}\rangle=0.7081|00\rangle+0.7013|11\rangle+0.0579|22\rangle+0.0579|33\rangle $\\
\cline{2-2}
4 & 
       $ A_{1}^1=\left(\begin{array}{l}
            \,\:\;0.4433 \\
            -0.8958 \\
            \,\:\;0.0232 \\
            \,\:\;0.0232
        \end{array}\right)$, $A_{2}^1=\left(\begin{array}{l}
            -0.3521 \\
            -0.9349 \\
            \,\:\;0.0317 \\
            \,\:\;0.0317
        \end{array}\right)$, $B_{1}^1=\left(\begin{array}{l}
            \,\:\;0.0553 \\
            -0.9940 \\
            \,\:\;0.0671 \\
            \,\:\;0.0671
        \end{array}\right)$,  $B_{2}^1=\left(\begin{array}{l}
            \,\:\;0.7673 \\
            -0.6400 \\
            \,\:\;0.0285 \\
            \,\:\;0.0285
        \end{array}\right)	$\\
\hline
 & 
$|\Psi_5^{22}\rangle=0.7071|11\rangle+0.7071|33\rangle $\\
\cline{2-2}
5 & 
        $A_{1}^1=\left(\begin{array}{l}
           \,\:\; 0\\
           \,\:\; 0.4366 \\
           \,\:\; 0\\
           \,\:\; 0.9000
        \end{array}\right)$, $A_{2}^1=\left(\begin{array}{l}
            \,\:\;0 \\
            -0.3275 \\
            \,\:\;0 \\
            \,\:\;0.9449
        \end{array}\right)$, $B_{1}^1=\left(\begin{array}{l}
           \,\:\; 0\\
           \,\:\; 0.0591 \\
           \,\:\; 0\\
           \,\:\; 0.9983
        \end{array}\right)$,  $B_{2}^1=\left(\begin{array}{l}
           \,\:\; 0\\
           \,\:\; 0.7476 \\
           \,\:\; 0\\
           \,\:\; 0.6641
        \end{array}\right)$\\
    \hline
\end{tabular}  
\caption{\label{I2222} Details of five entangled states and their corresponding measurements prepared for inequality $I_{22}^\text{F}$ in the experiment.}

\end{table}

\begin{table}[h]
    \centering
    \renewcommand\arraystretch{1.22}
    \scalebox{1}{
    \begin{tabular}{|c|c|}
\hline 
Order & Quantum state and its projection measurements for inequality  $I_{24}^\text{F}$\\
\hline 
1 & 
$|\Psi_1^{24}\rangle=0.6771|00\rangle+0.5006|11\rangle+0.5006|22\rangle+0.2008|33\rangle $\\
\hline
2 & $
|\Psi_2^{24}\rangle=0.6570|00\rangle+0.4858|11\rangle+0.4858|22\rangle+0.3104|33\rangle $\\
\hline
3 & $
|\Psi_3^{24}\rangle=0.6307|00\rangle+0.4663|11\rangle+0.4663|22\rangle+0.4089|33\rangle $\\
\hline
4 & $
|\Psi_4^{24}\rangle=0.6006|00\rangle+0.4440|11\rangle+0.4440|22\rangle+0.4949|33\rangle $\\
\hline
5 & $
|\Psi_5^{24}\rangle=0.5696|00\rangle+0.4204|11\rangle+0.4204|22\rangle+0.5696|33\rangle $\\
\hline
 &  $
\begin{array}{lll}
A_1^{1}=\left(\begin{array}{l}
           \,\:\; 0.5000 \\
           \,\:\; 0.4619-0.1913i \\
           \,\:\; 0.3536-0.3536i \\
           \,\:\; 0.1913-0.4619i
        \end{array}\right) & A_1^{2}=\left(\begin{array}{l}
             \,\:\;0.5000 \\
             \,\:\;0.1913+0.4619i \\
            -0.3536+0.3536i \\
            -0.4619-0.1913i 
        \end{array}\right) & A_1^{3}=\left(\begin{array}{l}
             \,\:\;0.5000       \\
            -0.4619+0.1913i \\
             \,\:\;0.3536-0.3536i \\
            -0.1913+0.4619i 
        \end{array}\right) \\
A_1^{4}=\left(\begin{array}{l}
             \,\:\;0.5000      \\
            -0.1913-0.4619i \\
            -0.3536+0.3536i \\
             \,\:\;0.4619+0.1913i 
        \end{array}\right) & A_2^{1}=\left(\begin{array}{l}
             \,\:\;0.5000      \\
             \,\:\;0.1913-0.4619i \\
            -0.3536-0.3536i \\
            -0.4619+0.1913i
        \end{array}\right) & A_2^{2}=\left(\begin{array}{l}
           \,\:\; 0.5000       \\
           \,\:\; 0.4619+0.1913i \\
           \,\:\; 0.3536+0.3536i \\
           \,\:\; 0.1913+0.4619i 
        \end{array}\right)\\
A_2^{3}=\left(\begin{array}{l}
             \,\:\;0.5000       \\
            -0.1913+0.4619i \\
            -0.3536-0.3536i \\
            \,\:\; 0.4619-0.1913i
        \end{array}\right) & A_2^{4}=\left(\begin{array}{l}
             \,\:\;0.5000       \\
            -0.4619-0.1913i \\
             \,\:\;0.3536+0.3536i \\
            -0.1913-0.4619i
        \end{array}\right) & B_1^{1}=\left(\begin{array}{l}
            \,\:\;0.5000       \\
             \,\:\;0.3536+0.3536i \\
            \,\:\;0.5000i     \\
             -0.3536+0.3536i 
        \end{array}\right) \\
B_1^{2}=\left(\begin{array}{l}
               \,\:\;0.5000 \\ 
            \,\:\;0.3536-0.3536 i \\
            -0.5000i     \\
            -0.3536-0.3536i 
        \end{array}\right) & B_1^{3}=\left(\begin{array}{l}
                \,\:\;0.5000      \\
              -0.3536-0.3536i \\
                \,\:\;0.5000i     \\
              \,\:\;0.3536-0.3536i
        \end{array}\right)  & B_1^{4}=\left(\begin{array}{l}
               \,\:\;0.5000    \\   
             -0.3536+0.3536i \\
                -0.5000i     \\
             \,\:\;0.3536+0.3536i 
        \end{array}\right) \\
B_2^{1}=\left(\begin{array}{l}
            \,\:\;0.5000         \\ 
            \,\:\;0.5000i        \\
           -0.5000 \\
            -0.5000i       
        \end{array}\right) & B_2^{2}=\left(\begin{array}{l}
           \,\:\; 0.5000 \\
           \,\:\; 0.5000 \\
           \,\:\; 0.5000 \\
           \,\:\; 0.5000
        \end{array}\right) & B_2^{3}=\left(\begin{array}{l}
                \,\:\;0.5000      \\    
                -0.5000i       \\ 
                 -0.5000 \\
                \,\:\;0.5000i       
        \end{array}\right)\\
B_2^{4}=\left(\begin{array}{l}
             \,\:\;0.5000i   \\     
            -0.5000\\
             \,\:\;0.5000\\
            -0.5000 
        \end{array}\right)&&
\end{array}  $\\
\cline{2-2}
 &  The measurements performed on the five quantum states are the same.\\
\hline
\end{tabular}  
}
\caption{\label{I2244} Details of five entangled states and their corresponding measurements prepared for inequality $I_{24}^\text{F}$ in the experiment.}
\end{table}

\begin{table}[h]
    \centering
    \renewcommand\arraystretch{1}
    \scalebox{1}{
    \begin{tabular}{|c|c|}
\hline 
Order & Quantum state and its projection measurements for inequality $I_{42}^\text{F}$ \\

\hline 
 & $
    |\Psi_1^{42}\rangle=0.4969|00\rangle+0.2949|11\rangle-0.5487|22\rangle+0.6042|33\rangle $\\
\cline{2-2}
1 & $
\begin{array}{llll}
A_{1}^1=\left(\begin{array}{l}
            -0.0656 \\
            -0.9763 \\
            -0.2025 \\
            \,\:\;0.0389
        \end{array}\right) & A_{2}^1=\left(\begin{array}{l}
            -0.8120 \\
            \,\:\;0.5109 \\
            \,\:\;0.2277 \\
            \,\:\;0.1665
        \end{array}\right) & A_{3}^1=\left(\begin{array}{l}
            -0.1680\\
            \,\:\;0.9640 \\
            -0.1947 \\
            \,\:\;0.0670
        \end{array}\right) & A_{4}^1=\left(\begin{array}{l}
            -0.6148 \\
            -0.6756 \\
            \,\:\;0.2171 \\
            \,\:\;0.3442
        \end{array}\right) \\
B_{1}^1=\left(\begin{array}{l}
            \,\:\;0.3578 \\
            -0.9295 \\
            \,\:\;0.0683 \\
            \,\:\;0.0577
        \end{array}\right) & B_{2}^1=\left(\begin{array}{l}
            \,\:\;0.2952  \\
            \,\:\;0.6395  \\
            -0.7041  \\
            \,\:\;0.0901
        \end{array}\right) & B_{3}^1=\left(\begin{array}{l}
           \,\:\; 0.1894 \\
           \,\:\; 0.9776 \\
           \,\:\; 0.0574 \\
           \,\:\; 0.0710
        \end{array}\right) & B_{4}^1=\left(\begin{array}{l}
            \,\:\;0.3064 \\
            -0.5990 \\
            -0.6889 \\
            \,\:\;0.2695
        \end{array}\right)
\end{array}$\\
\hline
 & $|\Psi_2^{42}\rangle=-0.4295|00\rangle-0.5566|11\rangle-0.6037|22\rangle+0.3760|33\rangle$\\
\cline{2-2}
2 & $
\begin{array}{llll}
A_{1}^1=\left(\begin{array}{l}
            \,\:\;0.1600  \\
            \,\:\;0.9245  \\
            -0.2162  \\
            \,\:\;0.2702 
        \end{array}\right) & A_{2}^1=\left(\begin{array}{l}
            \,\:\;0.3863  \\
            \,\:\;0.4204  \\
            -0.4466  \\
            \,\:\;0.6889 
        \end{array}\right) & A_{3}^1=\left(\begin{array}{l}
        -0.3785  \\
        \,\:\;0.1403  \\
        \,\:\;0.1751  \\
        \,\:\;0.8980 
        \end{array}\right) & A_{4}^1=\left(\begin{array}{l}
        -0.8092  \\
        -0.1995  \\
        -0.2435  \\
        \,\:\;0.4960 
        \end{array}\right) \\
B_{1}^1=\left(\begin{array}{l}
            -0.1870  \\
            -0.6240  \\
            \,\:\;0.1993  \\
            \,\:\;0.7321 
        \end{array}\right) & B_{2}^1=\left(\begin{array}{l}
        -0.0713  \\
        \,\:\;0.9654  \\
        -0.1135  \\
        \,\:\;0.2239 
        \end{array}\right) & B_{3}^1=\left(\begin{array}{l}
       \,\:\; 0.7378  \\
       \,\:\; 0.0489  \\
       \,\:\; 0.0841  \\
       \,\:\; 0.6680 
        \end{array}\right) & B_{4}^1=\left(\begin{array}{l}
        \,\:\;0.1464  \\
        -0.0655  \\
        -0.6832  \\
        \,\:\;0.7124 
        \end{array}\right)
\end{array}$\\
\hline
 & $|\Psi_3^{42}\rangle=0.2773|00\rangle-0.6556|11\rangle+0.2771|22\rangle+0.6454|33\rangle $\\
\cline{2-2}
3 & $
\begin{array}{llll}
A_{1}^1=\left(\begin{array}{l}
\,\:\;0.1936 \\
\,\:\;0.9136 \\
\,\:\;0.1936 \\
\,\:\;0.3007 \\
        \end{array}\right) & A_{2}^1=\left(\begin{array}{l}
       \,\:\; 0.6402 \\
\,\:\;0.2638 \\
\,\:\;0.6405 \\
\,\:\;0.3322 \\
        \end{array}\right) & A_{3}^1=\left(\begin{array}{l}
       \,\:\; 0.1208 \\
\,\:\;0.5424 \\
\,\:\;0.1205 \\
\,\:\;0.8226 \\
        \end{array}\right) & A_{4}^1=\left(\begin{array}{l}
       \,\:\; 0.6402 \\
\,\:\;0.2638 \\
\,\:\;0.6405 \\
\,\:\;0.3322 \\
        \end{array}\right) \\
B_{1}^1=\left(\begin{array}{l}
\,\:\;0.1399 \\
-0.7616 \\
\,\:\;0.1396 \\
\,\:\;0.6171 \\
        \end{array}\right) & B_{2}^1=\left(\begin{array}{l}
       \,\:\; 0.0431 \\
\,\:\;0.9611 \\
\,\:\;0.0433 \\
\,\:\;0.2695 \\
        \end{array}\right) & B_{3}^1=\left(\begin{array}{l}
        \,\:\;0.1399 \\
-0.7616 \\
\,\:\;0.1396 \\
\,\:\;0.6171 \\
        \end{array}\right) & B_{4}^1=\left(\begin{array}{l}
        -0.2636 \\
-0.1076 \\
-0.2645 \\
\,\:\;0.9214 \\
        \end{array}\right)
\end{array} $\\
\hline
 & $|\Psi_4^{42}\rangle=-0.2862|00\rangle-0.5980|11\rangle+0.5283|22\rangle+0.5304|33\rangle$\\
\cline{2-2}
4 & $
\begin{array}{llll}
A_{1}^1=\left(\begin{array}{l}
-0.4521 \\
-0.7375 \\
-0.1980 \\
\,\:\;0.4611 \\
        \end{array}\right) & A_{2}^1=\left(\begin{array}{l}
       \,\:\; 0.1358 \\
\,\:\;0.6595 \\
\,\:\;0.5079 \\
\,\:\;0.5373 \\
        \end{array}\right) & A_{3}^1=\left(\begin{array}{l}
        \,\:\;0.5679 \\
\,\:\;0.7373 \\
-0.3502 \\
\,\:\;0.1059 \\
        \end{array}\right) & A_{4}^1=\left(\begin{array}{l}
        \,\:\;0.2029 \\
-0.5501 \\
\,\:\;0.5728 \\
\,\:\;0.5728 \\
        \end{array}\right) \\
B_{1}^1=\left(\begin{array}{l}
-0.5320 \\
-0.7780 \\
\,\:\;0.2689 \\
\,\:\;0.1987 \\
        \end{array}\right) & B_{2}^1=\left(\begin{array}{l}
       \,\:\;0.5098 \\
\,\:\;0.2910 \\
-0.5665 \\
\,\:\;0.5783 \\
        \end{array}\right) & B_{3}^1=\left(\begin{array}{l}
       \,\:\; 0.1972 \\
\,\:\;0.8084 \\
\,\:\;0.3329 \\
\,\:\;0.4436 \\
        \end{array}\right) & B_{4}^1=\left(\begin{array}{l}
        -0.6361 \\
-0.3142 \\
-0.5318 \\
\,\:\;0.4624 \\
        \end{array}\right)
\end{array}$\\
\hline
 & $|\Psi_5^{42}\rangle=-0.5247|00\rangle+0.4473|11\rangle+0.4473|22\rangle+0.5485|33\rangle$\\
\cline{2-2}
5 & $
\begin{array}{llll}
A_{1}^1=\left(\begin{array}{l}
0.3180 \\
0.4095 \\
0.2417 \\
0.8202 \\
        \end{array}\right) & A_{2}^1=\left(\begin{array}{l}
        -0.5558 \\
\,\:\;0.5109 \\
\,\:\;0.6334 \\
\,\:\;0.1700 \\
        \end{array}\right) & A_{3}^1=\left(\begin{array}{l}
        -0.3487 \\
-0.1710 \\
\,\:\;0.0771 \\
\,\:\;0.9183 \\
        \end{array}\right) & A_{4}^1=\left(\begin{array}{l}
        \,\:\;0.4120 \\
-0.5906 \\
\,\:\;0.6322 \\
\,\:\;0.2858 \\
        \end{array}\right) \\
B_{1}^1=\left(\begin{array}{l}
0.2602 \\
0.3658 \\
0.2909 \\
0.8449 \\
        \end{array}\right) & B_{2}^1=\left(\begin{array}{l}
        -0.5302 \\
\,\:\;0.5978 \\
\,\:\;0.1185 \\
\,\:\;0.5894 \\
        \end{array}\right) & B_{3}^1=\left(\begin{array}{l}
        -0.1426 \\
-0.0761 \\
\,\:\;0.1686 \\
\,\:\;0.9723 \\
        \end{array}\right) & B_{4}^1=\left(\begin{array}{l}
        \,\:\;0.5572 \\
-0.3456 \\
-0.1358 \\
\,\:\;0.7427 \\
        \end{array}\right)
\end{array}$\\
 \hline
 & $|\Psi_6^{42}\rangle=-0.5904|00\rangle+0.5951|11\rangle-5360|22\rangle-0.0998|33\rangle$\\
\cline{2-2}
6 & $
\begin{array}{llll}
A_{1}^1=\left(\begin{array}{l}
\,\:\;0.9731 \\
-0.1052 \\
\,\:\;0.1511 \\
\,\:\;0.1381 \\
        \end{array}\right) & A_{2}^1=\left(\begin{array}{l}
        -0.5381 \\
\,\:\;0.7264 \\
\,\:\;0.4232 \\
\,\:\;0.0601 \\
        \end{array}\right) & A_{3}^1=\left(\begin{array}{l}
        \,\:\;0.7909 \\
\,\:\;0.1697 \\
-0.5566 \\
\,\:\;0.1892 \\
        \end{array}\right) & A_{4}^1=\left(\begin{array}{l}
        -0.6138 \\
-0.7744 \\
-0.1382 \\
\,\:\;0.0668 \\
        \end{array}\right) \\
B_{1}^1=\left(\begin{array}{l}
-0.8654 \\
-0.3458 \\
\,\:\;0.3494 \\
\,\:\;0.0970 \\
        \end{array}\right) & B_{2}^1=\left(\begin{array}{l}
        \,\:\;0.7047 \\
\,\:\;0.2670 \\
\,\:\;0.6371 \\
-0.1619 \\
        \end{array}\right) & B_{3}^1=\left(\begin{array}{l}
        -0.9031 \\
\,\:\;0.4136 \\
\,\:\;0.0674 \\
\,\:\;0.0941 \\
        \end{array}\right) & B_{4}^1=\left(\begin{array}{l}
        -0.3173 \\
\,\:\;0.3039 \\
\,\:\;0.8679 \\
\,\:\;0.2318 \\
        \end{array}\right)
\end{array}$\\
\hline
 & $|\Psi_7^{42}\rangle=0.7071|11\rangle+0.7071|33\rangle$\\
\cline{2-2}
7 & $
\begin{array}{llll}
A_{1}^1=\left(\begin{array}{l}
\,\:\;0 \\
\,\:\;0.5662 \\
\,\:\;0 \\
\,\:\;0.8243 \\
        \end{array}\right) & A_{2}^1=\left(\begin{array}{l}
       \,\:\; 0.7071 \\
\,\:\;0 \\
\,\:\;0.7071 \\
\,\:\;0 \\
        \end{array}\right) & A_{3}^1=\left(\begin{array}{l}
      \,\:\; 0\\
-0.1825 \\
\,\:\;0 \\
\,\:\;0.9832 \\
        \end{array}\right) & A_{4}^1=\left(\begin{array}{l}
       \,\:\; 0.7071 \\
\,\:\;0 \\
\,\:\;0.7071 \\
\,\:\;0 \\
        \end{array}\right) \\
B_{1}^1=\left(\begin{array}{l}
\,\:\;0 \\
\,\:\;0.2076 \\
\,\:\;0 \\
\,\:\;0.9782 \\
        \end{array}\right) & B_{2}^1=\left(\begin{array}{l}
       \,\:\; 0\\
\,\:\;0.8385 \\
\,\:\;0 \\
\,\:\;0.5449 \\
        \end{array}\right) & B_{3}^1=\left(\begin{array}{l}
       \,\:\; 0\\
\,\:\;0.2076 \\
\,\:\;0 \\
\,\:\;0.9782 \\
        \end{array}\right) & B_{4}^1=\left(\begin{array}{l}
        \,\:\;0 \\
-0.5449 \\
\,\:\;0 \\
\,\:\;0.8385 \\
        \end{array}\right)
\end{array}$\\
\hline
\end{tabular}  
}
\caption{\label{I4422} Details of seven entangled states and their corresponding measurements prepared for inequality $I_{42}^\text{F}$ in the experiment.}
\end{table}

\begin{table}[h]
    \centering
    \renewcommand\arraystretch{1.22}
    \begin{tabular}{|c|c|}
\hline 
Order & Quantum state and its projection measurements for inequality $I_{22}^\text{S}$ \\

\hline
1 & $
|\Phi_1^{22}\rangle=0.7071|00\rangle+0.7071|11\rangle$\\
\hline
2 & $
|\Phi_2^{22}\rangle=0.8406|00\rangle+0.5417|11\rangle$\\
\hline
3 & $
|\Phi_3^{22}\rangle=0.8861|00\rangle+0.4635|11\rangle$\\
\hline
 & $
\begin{array}{ll}
A_{1}^1=\left(\begin{array}{l}
\,\:\;0.7071 \\
\,\:\;0.5000-0.5000i \\
        \end{array}\right) & A_{1}^2=\left(\begin{array}{l}
        \,\:\;0.7071 \\
-0.5000+0.5000i \\
        \end{array}\right)  \\
A_{2}^1=\left(\begin{array}{l}
        \,\:\;0.7071 \\
-0.5000-0.5000i \\
        \end{array}\right) & A_{2}^2=\left(\begin{array}{l}
       \,\:\; 0.7071 \\
        \,\:\;0.5000+0.5000i \\
        \end{array}\right)\\ 
B_{1}^1=\left(\begin{array}{l}
        \,\:\;0.7071 \\
        \,\:\;0.7071i \\
        \end{array}\right) & B_{1}^2=\left(\begin{array}{l}
        \,\:\;0.7071 \\
        -0.7071i \\
        \end{array}\right)\\
B_{2}^1=\left(\begin{array}{l}
        \,\:\;0.7071 \\
        -0.7071 \\
        \end{array}\right) & B_{2}^2=\left(\begin{array}{l}
       \,\:\; 0.7071 \\
\,\:\;0.7071 \\
        \end{array}\right)
\end{array}$\\
\cline{2-2}
 & The measurements performed on the three quantum states are the same.\\
\hline
\end{tabular}
\caption{\label{I22} Details of three entangled states and their corresponding measurements prepared for inequality $I_{22}^\text{S}$ in the experiment.}
\end{table}

\begin{table}[h]
    \centering
    \renewcommand\arraystretch{1.22}
    \begin{tabular}{|c|c|}
\hline 
Order & Quantum state and its Settings for inequality $I_{32}^\text{S}$ \\
\hline
1 & $
|\Phi_1^{32}\rangle=0.7071|00\rangle+0.7071|11\rangle$\\
\hline
2 & $
|\Phi_2^{32}\rangle=0.8273|00\rangle+0.5618|11\rangle$\\
\hline
3 & $
|\Phi_3^{32}\rangle=0.8677|00\rangle+0.4971|11\rangle$\\
\hline
 & $
\begin{array}{ll}
A_{1}^1=\left(\begin{array}{l}
\,\:\;0.7071 \\
\,\:\;0.6124-0.3535i \\
        \end{array}\right) & A_{1}^2=\left(\begin{array}{l}
        \,\:\;0.7071 \\
-0.6124+0.3535i \\
        \end{array}\right) \\
A_{2}^1=\left(\begin{array}{l}
        \,\:\;0.7071 \\
-0.7071i \\
        \end{array}\right) & A_{2}^2=\left(\begin{array}{l}
       \,\:\; 0.7071 \\
\,\:\;0.7071i \\
        \end{array}\right) \\
A_{3}^1=\left(\begin{array}{l}
        \,\:\;0.7071 \\
-0.6124-0.3535i \\
        \end{array}\right) & A_{3}^2=\left(\begin{array}{l}
       \,\:\; 0.7071 \\
\,\:\;0.6124+0.3535i \\
        \end{array}\right) \\
B_{1}^1=\left(\begin{array}{l}
\,\:\;0.7071 \\
\,\:\;0.3535+0.6124i \\
        \end{array}\right) & B_{1}^2=\left(\begin{array}{l}
        \,\:\;0.7071 \\
-0.3535-0.6124i \\
        \end{array}\right) \\
B_{2}^1=\left(\begin{array}{l}
        \,\:\;0.7071 \\
-0.3535+0.6124i \\
        \end{array}\right) & B_{2}^2=\left(\begin{array}{l}
       \,\:\; 0.7071 \\
\,\:\;0.3535-0.6124i \\
        \end{array}\right) \\
B_{3}^1=\left(\begin{array}{l}
        \,\:\;0.7071 \\
-0.7071 \\
        \end{array}\right) & B_{3}^2=\left(\begin{array}{l}
       \,\:\; 0.7071 \\
\,\:\;0.7071 \\
        \end{array}\right)
\end{array}$\\
\cline{2-2}
 & The measurements performed on the three quantum states are the same.\\
\hline
\end{tabular}
\caption{\label{I32} Details of three entangled states and their corresponding measurements prepared for inequality $I_{32}^\text{S}$ in the experiment.}
\end{table}

\begin{table}[h]
    \centering
    \renewcommand\arraystretch{1.22}
    \begin{tabular}{|c|c|}
\hline 
Order & Quantum state and its Settings for inequality $I_{24}^\text{S}$ \\
\hline
1 & $
|\Phi_1^{24}\rangle=0.5000|00\rangle+0.5000|11\rangle+0.5000|22\rangle+0.5000|33\rangle$\\
\hline
2 & $
|\Phi_2^{24}\rangle=0.5505|00\rangle+0.5505|11\rangle+0.5505|22\rangle+0.3014|33\rangle$\\
\hline
3 & $
|\Phi_3^{24}\rangle=0.5635|00\rangle+0.5635|11\rangle+0.5635|22\rangle+0.2174|33\rangle$\\
\hline
 & $
\begin{array}{lll}
A_{1}^1=\left(\begin{array}{l}
\,\:\;0.5000 \\
\,\:\;0.4619-0.1913i \\
\,\:\;0.3535-0.3535i \\
\,\:\;0.1913-0.4619i \\
        \end{array}\right) & A_{1}^2=\left(\begin{array}{l}
        \,\:\;0.5000 \\
\,\:\;0.1913+0.4619i \\
-0.3535+0.3535i \\
-0.4619-0.1913i \\
        \end{array}\right) & A_{1}^3=\left(\begin{array}{l}
        \,\:\;0.5000 \\
-0.4619+0.1913i \\
\,\:\;0.3535-0.3535i \\
-0.1913+0.4619i \\
        \end{array}\right)\\
A_{1}^4=\left(\begin{array}{l}
      \,\:\; 0.5000 \\
-0.1913-0.4619i \\
-0.3535+0.3535i \\
\,\:\;0.4619+0.1913i \\
        \end{array}\right) & A_{2}^1=\left(\begin{array}{l}
\,\:\;0.5000 \\
\,\:\;0.1913-0.4619i \\
-0.3535-0.3535i \\
-0.4619+0.1913i \\
        \end{array}\right) & A_{2}^2=\left(\begin{array}{l}
       \,\:\; 0.5000 \\
\,\:\;0.4619+0.1913i \\
\,\:\;0.3535+0.3535i \\
\,\:\;0.1913+0.4619i \\
        \end{array}\right)\\
A_{2}^3=\left(\begin{array}{l}
        \,\:\;0.5000 \\
-0.1913+0.4619i \\
-0.3535-0.3535i \\
\,\:\;0.4619-0.1913i \\
        \end{array}\right) & A_{2}^4=\left(\begin{array}{l}
        \,\:\;0.5000 \\
-0.4619-0.1913i \\
\,\:\;0.3535+0.3535i \\
-0.1913-0.4619i \\
        \end{array}\right) & B_{1}^1=\left(\begin{array}{l}
\,\:\;0.5000 \\
\,\:\;0.3535+0.3535i \\
\,\:\;0.5000i \\
-0.3535+0.3535i \\
        \end{array}\right) \\
B_{1}^2=\left(\begin{array}{l}
      \,\:\; 0.5000 \\
\,\:\;0.3535-0.3535i \\
-0.5000i \\
-0.3535-0.3535i \\
        \end{array}\right) & B_{1}^3=\left(\begin{array}{l}
        \,\:\;0.5000 \\
-0.3535-0.3535i \\
\,\:\;0.5000i \\
\,\:\;0.3535-0.3535i \\
        \end{array}\right) & B_{1}^4=\left(\begin{array}{l}
        \,\:\;0.5000 \\
-0.3535+0.3535i \\
-0.5000i \\
\,\:\;0.3535+0.3535i \\
        \end{array}\right)\\ 
B_{2}^1=\left(\begin{array}{l}
\,\:\;0.5000 \\
\,\:\;0.5000i \\
-0.5000 \\
-0.5000i \\
        \end{array}\right) & B_{2}^2=\left(\begin{array}{l}
       \,\:\; 0.5000 \\
\,\:\;0.5000 \\
\,\:\;0.5000\\
\,\:\;0.5000 \\
        \end{array}\right) & B_{2}^3=\left(\begin{array}{l}
        \,\:\;0.5000 \\
-0.5000i \\
-0.5000 \\
\,\:\;0.5000i \\
        \end{array}\right) \\
B_{2}^4=\left(\begin{array}{l}
        \,\:\;0.5000 \\
-0.5000 \\
\,\:\;0.5000 \\
-0.5000 \\
        \end{array}\right) &&
\end{array}$\\
\cline{2-2}
 & The measurements performed on the three quantum states are the same.\\
\hline
\end{tabular}
\caption{\label{I24} Details of three entangled states and their corresponding measurements prepared for inequality $I_{24}^\text{S}$ in the experiment.}
\end{table}

\begin{table}[h]
    \centering
    \renewcommand\arraystretch{1.22}
    \begin{tabular}{|c|c|}
\hline 
Order & Quantum state and its Settings for inequality $I_{34}^\text{S}$ \\
\hline
1 & $
|\Phi_1^{34}\rangle=0.5000|00\rangle+0.5000|11\rangle+0.5000|22\rangle+0.5000|33\rangle$\\
\hline
2 & $
|\Phi_2^{34}\rangle=0.5452|00\rangle+0.5452|11\rangle+0.5452|22\rangle+0.3293|33\rangle$\\
\hline
3 & $
|\Phi_3^{34}\rangle=0.5588|00\rangle+0.5588|11\rangle+0.5588|22\rangle+0.2517|33\rangle$\\
\hline
 & $
\begin{array}{lll}
A_1^{1}=\left(\begin{array}{l}
\,\:\;0.5000 \\
\,\:\;0.4830-0.1294i \\
\,\:\;0.4330-0.2500i \\
\,\:\;0.3535-0.3535i \\
        \end{array}\right) & A_1^{2}=\left(\begin{array}{l}
        \,\:\;0.5000 \\
\,\:\;0.1294+0.4830i \\
-0.4330+0.2500i \\
-0.3535-0.3535i \\
        \end{array}\right) & A_1^{3}=\left(\begin{array}{l}
        \,\:\;0.5000 \\
-0.4830+0.1294i \\
\,\:\;0.4330-0.25i \\
-0.3535+0.3535i \\
        \end{array}\right) \\
A_1^{4}=\left(\begin{array}{l}
        \,\:\;0.5000 \\
-0.1294-0.4830i \\
-0.4330+0.2500i \\
\,\:\;0.3535+0.3535i \\
        \end{array}\right) & A_2^{1}=\left(\begin{array}{l}
\,\:\;0.5000 \\
\,\:\;0.3535-0.3535i \\
-0.5000i \\
-0.3535-0.3535i \\
        \end{array}\right) & A_2^{2}=\left(\begin{array}{l}\
       \,\:\; 0.5000 \\
\,\:\;0.3535+0.3535i \\
\,\:\;0.5000i \\
-0.3535+0.3535i \\
        \end{array}\right)\\ 
A_2^{3}=\left(\begin{array}{l}
       \,\:\; 0.5000 \\
-0.3535+0.3535i \\
-0.5000i \\
\,\:\;0.3535+0.3535i \\
        \end{array}\right) & A_2^{4}=\left(\begin{array}{l}
        \,\:\;0.5000 \\
-0.3535-0.3535i \\
\,\:\;0.5000i \\
\,\:\;0.3535-0.3535i \\
        \end{array}\right) & A_3^{1}=\left(\begin{array}{l}
\,\:\;0.5000 \\
\,\:\;0.1294-0.4830i \\
-0.4330-0.2500i \\
-0.3535+0.3535i \\
        \end{array}\right) \\
A_3^{2}=\left(\begin{array}{l}
       \,\:\; 0.5000 \\
\,\:\;0.483+0.1294i \\
\,\:\;0.433+0.25i \\
\,\:\;0.3535+0.3535i \\
        \end{array}\right) & A_3^{3}=\left(\begin{array}{l}
       \,\:\;0.5000 \\
-0.1294+0.4830i \\
-0.4330-0.2500i \\
\,\:\;0.3535-0.3535i \\
        \end{array}\right) & A_3^{4}=\left(\begin{array}{l}
        \,\:\;0.5000 \\
-0.4830-0.1294i \\
\,\:\;0.4330+0.2500i \\
-0.3535-0.3535i \\
        \end{array}\right)\\ 
B_1^{1}=\left(\begin{array}{l}
\,\:\;0.5000 \\
\,\:\;0.433+0.2500i \\
\,\:\;0.2500+0.4330i \\
\,\:\;0.5000i \\
        \end{array}\right) & B_1^{2}=\left(\begin{array}{l}
        \,\:\;0.5000 \\
\,\:\;0.2500-0.4330i \\
-0.2500-0.4330i \\
-0.5000 \\
        \end{array}\right) & B_1^{3}=\left(\begin{array}{l}
        \,\:\;0.5000 \\
-0.4330-0.2500i \\
\,\:\;0.2500+0.4330i \\
-0.5000i \\
        \end{array}\right) \\
B_1^{4}=\left(\begin{array}{l}
        \,\:\;0.5000 \\
-0.2500+0.4330i \\
-0.2500-0.4330i \\
\,\:\;0.5000 \\
        \end{array}\right)  & B_2^{1}=\left(\begin{array}{l}
\,\:\;0.5000 \\
\,\:\;0.2500+0.433i \\
-0.25+0.433i \\
-0.5000 \\
        \end{array}\right) & B_2^{2}=\left(\begin{array}{l}
        \,\:\;0.5000 \\
\,\:\;0.4330-0.2500i \\
\,\:\;0.2500-0.4330i \\
-0.5000i \\
        \end{array}\right) \\
B_2^{3}=\left(\begin{array}{l}
      \,\:\; 0.5000 \\
-0.2500-0.4330i \\
-0.2500+0.4330i \\
\,\:\;0.5000 \\
        \end{array}\right) & B_2^{4}=\left(\begin{array}{l}
        \,\:\;0.5000 \\
-0.4330+0.2500i \\
\,\:\;0.2500-0.4330i \\
\,\:\;0.5000i \\
        \end{array}\right)  & B_3^{1}=\left(\begin{array}{l}
\,\:\;0.5000 \\
\,\:\;0.5000i \\
-0.5000 \\
-0.5000i \\
        \end{array}\right) \\
B_3^{2}=\left(\begin{array}{l}
       \,\:\; 0.5000 \\
\,\:\;0.5000 \\
\,\:\;0.5000 \\
\,\:\;0.5000 \\
        \end{array}\right) & B_3^{3}=\left(\begin{array}{l}
        \,\:\;0.5000 \\
-0.5000i \\
-0.5000 \\
\,\:\;0.5000i \\
        \end{array}\right) & B_3^{4}=\left(\begin{array}{l}
        \,\:\;0.5000 \\
-0.5000 \\
\,\:\;0.5000 \\
-0.5000 \\
        \end{array}\right)  \\
\end{array}$\\
\cline{2-2}
 & The measurements performed on the three quantum states are the same.\\
\hline
\end{tabular}
\caption{\label{I34} Details of three entangled states and their corresponding measurements prepared for inequality $I_{34}^\text{S}$ in the experiment.}
\end{table}

\clearpage

\bibliography{Ref_PRL_round1}